\documentclass[10pt,twocolumn]{article}
\usepackage[top=1.9cm, bottom=1.9cm, left=1.7cm, right=1.7cm]{geometry}
\usepackage{times}
\usepackage[hyphens]{url}  %
\usepackage[keeplastbox]{flushend}  %
\usepackage{graphicx}  %

\makeatletter
\def\url@leostyle{%
  \@ifundefined{selectfont}{\def\UrlFont{\small}}%
  {\def\UrlFont{}}%
}
\makeatother
\usepackage{xcolor}
\usepackage{color}
\usepackage[small,bf]{caption}

\usepackage{paralist}
\usepackage{subfigure}
\usepackage{booktabs}
\newcommand{\descr}[1]{\smallskip\noindent\textbf{#1}}
\usepackage{multirow}
\usepackage{flushend}
\usepackage{enumitem}

\usepackage{algorithm}
\usepackage{algpseudocode} 
\AtBeginDocument{%
  \providecommand\BibTeX{{%
    \normalfont B\kern-0.5em{\scshape i\kern-0.25em b}\kern-0.8em\TeX}}}

\usepackage[skip=2pt]{caption}
\definecolor{linkcol}{rgb}{0,0,0.5}
\definecolor{citecol}{rgb}{0,0.5,0.3}
\definecolor{urlcol}{rgb}{0.3,0,0}

\usepackage[bookmarks=true,%
            bookmarksnumbered=true,%
            colorlinks=true,%
            linkcolor=linkcol,%
            citecolor=citecol,%
            urlcolor=urlcol,%
            hypertexnames=true,%
            pdfpagelabels,
      ]%
      {hyperref}

\usepackage[OT2,OT1]{fontenc}
\newcommand\cyr
{
\renewcommand\rmdefault{wncyr}
\renewcommand\sfdefault{wncyss}
\renewcommand\encodingdefault{OT2}
\normalfont
\selectfont
}
\DeclareTextFontCommand{\textcyr}{\cyr}

\usepackage[compact]{titlesec}
\titlespacing*{\section}{0pt}{*2}{2pt}
\titlespacing{\subsection}{0pt}{*2}{2pt}

\newif\ifcomment
\commenttrue
\ifcomment
\newcommand{\sz}[1]{{\bf \textcolor{blue}{SZ: #1}}}
\newcommand{\kp}[1]{{\textcolor{orange}{#1}}}
\newcommand{\kpnote}[1]{{\bf \textcolor{green}{KP: #1}}}
\newcommand{\gs}[1]{{\bf \textcolor{red}{GS: #1}}}
\newcommand{\edc}[1]{{\bf \textcolor{brown}{EDC: #1}}}
\newcommand{\ms}[1]{{\bf \textcolor{blue}{MS: #1}}}
\newcommand{\jbnote}[1]{{\bf \textcolor{blue}{JB: #1}}}
\newcommand{\revision}[1]{{\textcolor{black}{#1}}}
\newcommand{\revcomment}[1]{{\textcolor{red}{}}}
\newcommand{\revisionfinal}[1]{{\textcolor{black}{#1}}}
\newcommand{\hide}[1]{#1}
\else
\newcommand{\sz}[1]{}
\newcommand{\kp}[1]{}
\newcommand{\kpnote}[1]{}
\newcommand{\gs}[1]{}
\newcommand{\edc}[1]{}
\newcommand{\ms}[1]{}
\newcommand{\jbnote}[1]{}
\newcommand{\revision}[1]{}
\newcommand{\revcomment}[1]{}
\newcommand{\revisionfinal}[1]{}
\newcommand{\hide}[1]{}
\fi

\begin{document}

\title{\bf ``It is just a flu'': Assessing the Effect of Watch History on YouTube's Pseudoscientific Video Recommendations\thanks{To appear at the 16th International Conference on Web and Social Media (ICWSM 2022). Please cite the ICWSM version.}}

\author{Kostantinos Papadamou$^\star$, Savvas Zannettou$^\mp$, Jeremy Blackburn$^\dagger$\\
Emiliano De Cristofaro$^\ddagger$, Gianluca Stringhini$^\diamond$, Michael Sirivianos$^\star$\\[0.5ex]
\normalsize $^\star$Cyprus University of Technology, $^\mp$Max Planck Institute, $^\dagger$Binghamton University \\[-0.5ex]
\normalsize $^\ddagger$University College London, $^\diamond$Boston University\\
\normalsize ck.papadamou@edu.cut.ac.cy, szannett@mpi\-inf.mpg.de, jblackbu@binghamton.edu \\[-0.5ex]
\normalsize e.decristofaro@ucl.ac.uk, gian@bu.edu, michael.sirivianos@cut.ac.cy
}
\date{}

\maketitle

\begin{abstract}
The role played by YouTube's recommendation algorithm in unwittingly promoting misinformation and conspiracy theories is not entirely understood. %
Yet, this can have dire real-world consequences, especially when pseudoscientific content is promoted to users at critical times, such as the COVID-19 pandemic.
In this paper, we set out to characterize and detect pseudoscientific misinformation on YouTube.
We collect 6.6K videos related to COVID-19, the Flat Earth theory, as well as the anti-vaccination and anti-mask movements. 
Using crowdsourcing, we annotate them as pseudoscience, legitimate science, or irrelevant and train a deep learning classifier to detect pseudoscientific videos with an accuracy of $0.79$.

We quantify user exposure to this content on various parts of the platform 
and how this exposure changes based on the user's watch history.
We find that YouTube suggests more pseudoscientific content regarding traditional pseudoscientific topics (e.g., flat earth, anti-vaccination) than for emerging ones (like COVID-19). 
At the same time, these recommendations are more common on the search results page than on a user's homepage or in the recommendation section when actively watching videos. %
Finally, we shed light on how a user's watch history substantially affects the type of recommended videos.
\end{abstract}

\section{Introduction}
\label{sec:introduction}
User-generated video platforms like YouTube have exploded in popularity over the last decade. %
For many users, it has also become one of the most important information sources for news and various other topics~\cite{reuters2020digitalnewsreport}. %
Alas, such platforms are also often fertile ground for the spread of misleading and potentially harmful information like conspiracy theories and health-related disinformation~\cite{comsos2019youtubeconspiracy}.%

YouTube and other social media platforms have struggled with mitigating the harm from this type of content. The difficulty is partly due to the sheer scale and also because of the deployment of recommendation algorithms~\cite{fastcompany2019youtubeconspiracies}.
Purely automated moderation tools have thus far been insufficient to moderate content, and human moderators had to be brought back into the loop~\cite{vincent2020YouTubeBringsBack}.
Additionally, the machine learning algorithms that YouTube relies on to recommend content to users also recommend potentially harmful content~\cite{papadamou2020disturbed}, and their opaque nature makes them difficult to audit.

For certain types of content, e.g., health-related topics, harmful videos can have devastating effects on society, especially during crises like the COVID-19 pandemic~\cite{bbc2020youtubecoronavirus}.
For instance, %
conspiracy theories have suggested that COVID-19 is caused by 5G~\cite{covid5g2020conspiracy} or Bill Gates~\cite{covid2020billgates}, hindering social distancing, masking, and vaccination efforts%
~\cite{enserink2020fact}. %
Conspiracy theories are usually built on tenuous connections between various events, with little to no actual evidence to support them. 
On user-generated content platforms like YouTube, these are often presented as facts, regardless of whether they are supported by facts and even though they have been widely debunked.
\revisionfinal{
Motivated by the pressing need to mitigate the spread of pseudoscientific content, we focus on detecting and characterizing pseudoscientific and conspiratorial content on YouTube, while assessing the effect of a user's watch history on YouTube's pseudoscientific video recommendations.
}
In particular, we aim to: 1) assess how likely it is for users with distinct watch histories to come across pseudoscientific content on YouTube, and 2) analyze how YouTube's recommendation algorithm contributes to promoting pseudoscience.
To do so, we set out to answer the following two research questions:
\begin{compactenum}

\item[\bf RQ1] Can we effectively detect and characterize pseudoscientific content on YouTube?
\item[\bf RQ2] What is the proportion of pseudoscientific content on the homepage of a YouTube user, in search results, and the video recommendations section of YouTube? How are these proportions affected by the user's watch history? 

\end{compactenum}

\descr{Methodology.} 
We focus on four pseudoscientific topics: 1)~COVID-19, 2)~Flat Earth theory, 3)~anti-vaccination, and 4)~anti-mask movement.
We collect 6.6K unique videos and use crowdsourcing to label them in three categories: science, pseudoscience, or irrelevant. %
We then assign labels to each video based on the majority agreement of the annotators. 
We excluded videos where all the annotators disagreed resulting in a final dataset of 5.7K videos.
Using this dataset, we train a deep learning classifier to detect pseudoscientific content across multiple topics on YouTube.

\revisionfinal{
Next, the challenge we face in assessing the effect of watch history on YouTube recommendations lies in faithfully recreating the behavior of real users with particular profiles and interests.
In addressing similar questions, Hussein et al.~\cite{hussein2020measuring} create various user profiles with distinct demographics and watch history and they perform search queries to investigate the effects of these personalization attributes on the amount of misinformation in YouTube search results.
Their crawler watches a curated subset of the videos returned by the search queries to build the watch history of their user profiles. At the same time, it collects the \emph{Up-Next} and the \emph{Top-5} video recommendations to assess how the profile affects the videos listed on the recommendations section of the platform.
}

\revisionfinal{
Taking cues from this approach, in this work we set out to emulate a real-user's behavior on the platform (see Section~\ref{subsec:experiments_details}), while focusing on the effect of the user's watch history on various parts of the platform, including the user's homepage.
Compared to Hussein et al.~\cite{hussein2020measuring}, we perform a more comprehensive
measurement of what a user who follows YouTube's recommendations encounters.
}
\revisionfinal{
The differences in our methodology lead to some interesting complementary results (see Section~\ref{subsec:experiments_take_aways}).
}
\revcomment{(Comment: MT1)}

We perform our experiments using three carefully crafted user profiles, each with a different watch history, while all other account information remains the same, to emulate logged-in users.
We also perform a set of experiments using a browser without a Google account to emulate non-logged-in users and another set using the YouTube Data API exclusively.
To populate the watch history of the three user profiles, we devise a methodology to identify the minimum amount of videos that must be watched by a user before YouTube's recommendation algorithm starts generating substantially personalized recommendations.
We build three distinct profiles:
1) a user interested in scientific content;
2) a user interested in pseudoscientific content; and
3) a user interested in both scientific and pseudoscientific content.
Using these profiles, we perform three experiments to quantify the user's exposure to pseudoscientific content on various parts of the platform and how this exposure changes based on a user's watch history. 
Note that we manually review all the videos classified as pseudoscientific in all experiments.
\descr{Findings.}
Overall, our study leads to the following findings:
\begin{compactenum}
\item We can detect pseudoscientific content, as our deep learning classifier yields $0.79$ accuracy and outperforms SVM, Random Forest, and BERT-based classifiers \textbf{(RQ1)}. %

\item We find that the minimum amount of videos a user needs to watch before YouTube learns her interests and starts generating more personalized science and pseudoscience-related recommendations is 22 \textbf{(RQ2)}.

\item The watch history of the user substantially affects search results and related video recommendations. At the same time, pseudoscientific videos are more likely to appear in search results than in the video recommendations section or the user's homepage \textbf{(RQ2)}.

\item In traditional pseudoscience topics (e.g., Flat Earth), there is a higher rate of recommended pseudoscientific content than in more recent issues like COVID-19, anti-vaccination, and anti-mask. 
For COVID-19, we find an even smaller amount of pseudoscientific content being suggested. This indicates that YouTube took partly effective measures to mitigate pseudoscientific misinformation related to the COVID-19 pandemic \textbf{(RQ2)}.

\item The YouTube Data API results are similar to those of the non-logged-in profile with no watch history (using a browser); this indicates that recommendations returned using the API are not subject to personalization.

\end{compactenum}

\descr{Contributions.} 
To the best of our knowledge, we present the first study focusing on multiple health-related pseudoscientific topics on YouTube pertaining to the COVID-19 pandemic.
\revisionfinal{
We develop a complete and reusable framework that allows us to assess the prevalence of pseudoscientific content on various parts of the YouTube platform (i.e., homepage, search results, video recommendations) while accounting for the effect of a user's watch history.
}
Our methodology and software tools can be re-used for other studies focusing on other topics of interest.
We also publish our ground-truth dataset, the classifier, the source code/crawlers used in our experiments, and the source code of our framework. 
We make publicly available our ground-truth dataset\footnote{\url{https://zenodo.org/record/4769731}}, the classifier, the source code/crawlers used in our experiments, and the source code of our framework\footnote{\url{https://github.com/kostantinos-papadamou/pseudoscience-paper}}. 
We are confident that this will help the research community shed additional light on YouTube's recommendation algorithm and its potential influence. 
\section{Dataset \& Annotation}
\label{sec:methodology}
In this section, we present our data collection and crowdsourced annotation methodology.

\subsection{Data Collection}
\label{subsec:data_collection}

Since we aim to detect pseudoscientific video content automatically, we collect a set of YouTube videos related to four, arguably relevant, topics: %
1)~COVID-19~\cite{buzzfeedcoronavirus_2020}, 
2)~the anti-vaccine movement~\cite{antiVaxx2020}, 
3)~the anti-mask movement~\cite{antiMask2020}, and 
4)~the Flat Earth theory~\cite{guardianflatearth_2019}.
We focus on COVID-19 and the anti-mask movement because both are timely topics of great societal interest.
We also choose anti-vaccination because it is both an increasingly popular and traditional pseudoscientific topic. Last, we include the Flat Earth theory because it is a ``long-standing'' pseudoscientific subject.

\begin{table}[t!]
\centering
 \setlength{\tabcolsep}{3pt}
\small
\begin{tabular}{lrrrr}
\toprule
\textbf{Pseudoscientific Topic} & \textbf{\#Seed} & \textbf{\#Recommended}  \\
\midrule
COVID-19 & 378 & 1,645 \\
Anti-vaccination & 346 & 1,759 \\
Anti-mask & 199 & 912 \\
Flat Earth & 200 & 1,211 \\
\midrule
\textbf{Total} & \textbf{1,123} & \textbf{5,527} \\
\bottomrule
\end{tabular}%
\caption{Overview of the collected data: number of seed videos and number of their recommended videos.}
\label{tab:dataset_overview}
\end{table}

Then, for each topic of interest, we define search queries and use them to search YouTube and collect videos.
For COVID-19 we search using the terms ``COVID-19'' and ``coronavirus,'' and for the anti-vaccination movement we use the terms ``anti-vaccination'' and ``anti-vaxx''. 
On the other hand, for the anti-mask movement and the Flat Earth theory we only use the terms ``anti-mask'' and ``flat earth,'' respectively, since there are no other terms that point to the same definition as is the case for the 
other two topics.
Next, we search YouTube using the YouTube Data API~\cite{youtubedataapi_2020} and the search queries defined for each topic.
For each search query of each selected topic we obtain the first 200 videos as returned by the API's search functionality. 
We refer to those videos as the ``seed'' videos of our data collection methodology.
Additionally, for each seed video, we collect the top 10 recommended videos associated with it, as returned by the API.
We perform our data collection on August 1-20, 2020, collecting 6.6K unique videos (1.1K seed videos and 5.5K videos recommended from the seed videos).
Table~\ref{tab:dataset_overview} summarizes our dataset.

For each video in our dataset, we collect:
1) the video title and description;
2) a set of tags defined by the uploader;
3) transcript;
4) video statistics (e.g., the number of views, likes, etc.); and
5) the 200 top comments, defined by YouTube's relevance metric, without their replies.

\subsection{Crowdsourcing Data Annotation}
\label{subsec:manual_annotation}
To create a ground-truth dataset of scientific and pseudoscientific videos, we use the Appen platform~\cite{appen_2020} to get crowdsourced annotations for all the collected videos. 
We present each video to three annotators who inspect its content and metadata to assign one of three labels:

\begin{compactenum}
\item \descr{Science.} The content is related to any scientific field that systematically studies the natural world's structure and the behavior or humanity's artifacts (e.g., Chemistry, Biology, Mathematics, Computer Science, etc.).
Videos that debunk science-related conspiracy theories (e.g., explaining why 5G technology is not harmful) also fall in this category.
For example, a COVID-19 video with an expert estimating the total number of cases or excess deaths falls in this category if the estimation rests on the scientific consensus and official data.
\item \descr{Pseudoscience.} %
The video meets at least one of the following criteria:
a) holds a view of the world that goes against the scientific consensus (e.g., anti-vaccine movement);
b) comprises statements or beliefs that are self-fulfilling or unfalsifiable (e.g., Meditation); %
c) develops hypotheses that are not evaluated following the scientific method (e.g., Astrology); or
d) explains events as secret plots by powerful forces rather than overt activities or accidents (e.g., the 5G-coronavirus conspiracy theory).
\item \descr{Irrelevant.} %
The content is not relevant to any scientific field and does not fall in the Pseudoscience category. 
For example, music videos and cartoon videos are considered irrelevant. Conspiracy theory debunking videos that are not relevant to a scientific field are deemed irrelevant (e.g., a video debunking the Pizzagate conspiracy theory).
\end{compactenum}

\descr{Annotation.}
The annotation process is carried out by 992 annotators, both male and female, recruited through the Appen platform. We give annotators instructions on what constitutes scientific and pseudoscientific content using appropriate descriptions and several examples. They are offered $\$0.03$ per annotation. 
Three annotators label each video. 
To ease the annotation process, we provide a clear description of the task and our labels, and all video information that an annotator needs to inspect and correctly annotate a video.
Screenshots of the instructions are available, anonymously, from~\cite{anonymouspaperresources}.
Appen provides no demographic information about the annotators, other than an assurance that they are experienced and attained high accuracy in other tasks.
To assess the annotators' quality, before allowing them to submit annotations, we ask them to annotate 5 test videos randomly selected from a set of 54 test videos annotated by the first author of this paper.
An annotator can submit annotations only when she labels at least 3 out of the 5 test videos correctly.
This initial test guarantees that our annotators are more likely to have a scientific rather than conspiratorial pseudoscientific outlook, which would probably pollute our results.

\begin{table}[t!]
\centering
\small
 \setlength{\tabcolsep}{3pt}
\begin{tabular}{lrrr}
\toprule
\textbf{Topic} & \textbf{\#Pseudoscience} & \textbf{\#Other}  \\
\midrule
COVID-19 & 368 & 1,328 \\
Anti-vaccination & 394 & 1,423 \\
Anti-mask & 188 & 789 \\
Flat Earth & 375 & 869 \\
\midrule
\textbf{Total} & \textbf{1,325} & \textbf{4,409} \\
\bottomrule
\end{tabular}%
\caption{Overview of our final ground-truth dataset.}
\label{tab:final_groundtruth_dataset_overview}
\end{table}

Furthermore, using the collected annotations, we calculate the Fleiss' Kappa Score ($k$)~\cite{fleiss1971measuring} to assess the annotators' agreement.
We get $k=0.14$, which is considered ``slight'' agreement.
To mitigate the effect of the low agreement score on our results, we first exclude from our dataset all the 915 videos ($13.8\%$) where all annotators disagreed with each other and we calculate again the agreement score.
We get $k=0.24$, which is considered ``fair'' agreement.
Next, we assign labels to each video in our ground-truth dataset based to the majority agreement of all the annotators resulting in a ground-truth dataset that includes 1,197 science, 1,325 pseudoscience, and 3,212 irrelevant videos. %
Last, to further mitigate the effects of the low agreement score of our crowdsourced annotation, we collapse our three labels into two, combining the science with the irrelevant videos into an ``Other'' category.
This yields a final ground-truth dataset with 1,325 pseudoscience and 4,409 other videos (see Table~\ref{tab:final_groundtruth_dataset_overview}).

\descr{Performance Evaluation.}
To evaluate our crowdsourced annotation performance, we randomly select 600 videos from our ground-truth dataset and manually annotate them.
Using the first author's annotations as ground-truth, we calculate the precision, recall, and F1 score of our crowdsourced annotation, yielding respectively $0.92$, $0.91$, and $0.92$.
We argue that this represents an acceptable performance given the subjective nature of scientific and pseudoscientific content.
\subsection{Ethics}
In this work, we only collect publicly available data and we make no attempt to de-anonymize users. %
Overall, we follow standard ethical guidelines~\cite{dittrich2012menlo,rivers2014ethical} regarding information research and the use of shared measurement data.
More precisely, we ensure compliance with GDPR's~\cite{gdpr2018eu} ``Right to be Forgotten'' and ``Right of Access'' principles.
We have also obtained ethics approval from the first author's national ethics committee to ensure that our crowdsourced annotation process does not pose risks to the annotators. 
Nevertheless, we consider the detrimental effects of the controversial content we study.
For this reason we inform our annotators and enable them to opt-out our annotation process at any time.
Finally, we acknowledge that the price offered per annotation is quite low and this is mainly because of the large number of videos we needed to annotate.
However, this price allows us to acquire all the required annotations within the budget allocated for this research.

\begin{figure*}[t!]
\centering
\includegraphics[width=0.6\linewidth]{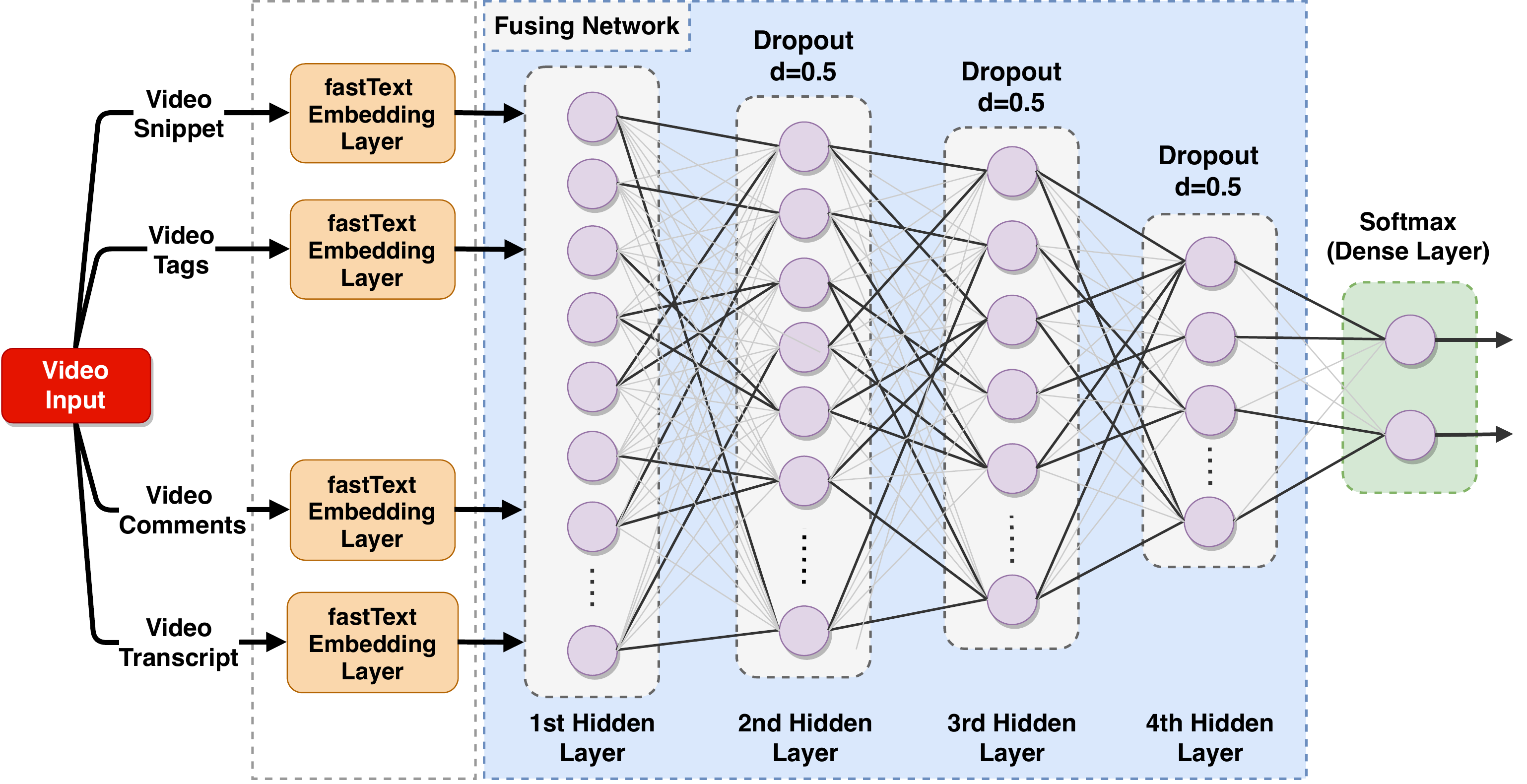}
\caption{Architecture of our deep learning classifier for the detection of pseudoscientific videos.}
\label{fig:model_architecture}
\end{figure*}

\section{Detection of Pseudoscientific Videos (RQ1)}
\label{sec:pseudoscience_detection}

In this section, we present our classifier geared towards detecting pseudoscientific videos.
To train and test it, we use our ground-truth dataset of 5,734 (1,325 pseudoscience and 4,409 other) videos. 

\subsection{Classifier Architecture}
\label{model_architecture}
Figure~\ref{fig:model_architecture} depicts the architecture of our classifier. 
The classifier consists of four different branches, each processing a distinct input feature type: snippet, video tags, transcript, and the top 200 comments of a video. 
Then, all four branches' outputs are concatenated to form a five-layer, fully-connected neural network that merges their output and drives the final classification.
We choose to build a classifier that analyzes the textual metadata (e.g., transcript) of a video and the discussions associated with it (i.e., comments) because we believe that they can provide a more meaningful signal about the pseudoscientific stance of a video than other types of input (i.e., thumbnail).
Also, the transcript of the video allows us to also consider the main themes discussed in the actual video by the creator/uploader of the video.
The classifier uses fastText~\cite{fasttext_2020}, a library %
for efficient learning of word/document-level vector representations and sentence classification, to generate vector representations (embeddings) for all the available video metadata in text.
For each type of input feature, we use the pre-trained fastText embeddings released in~\cite{mikolov2018advances} and fine-tune them for our text classification task using each of our input features.
These fine-tuned models extract a 300-dimensional vector representation for each of the following input features of our dataset:
\begin{compactitem}
\item {\em Snippet.} 
Concatenation of a video's title and description. 

\item {\em Tags.}
Words defined by the uploader of a video to describe the content of the video.

\item  {\em Transcript.}
Naturally, this is one of the most important features, as it describes the video's actual content.
(It includes the subtitles uploaded by the creator of the video or auto-generated by YouTube.) 
The classifier uses the fine-tuned model to learn a vector representation of the concatenated text of the transcript.

\item {\em Comments.}
We consider the top 200 comments of the video as returned by the YouTube Data API.
We first concatenate each video's comments and use them to fine-tune the fastText model and extract vector representations.
\end{compactitem}

\smallskip The second part of the classifier (the ``Fusing Network'' in Figure~\ref{fig:model_architecture}) is essentially a four-layer, fully-connected, dense neural network. 
We use a Flatten utility layer to merge the outputs of the four branches of the first part of the classifier, creating a 1200-dimensional vector.
This vector is processed by the four subsequent layers comprising 256, 128, 64, and 32 units, respectively, with ReLU activation.
To avoid overfitting, we regularize using the Dropout technique; %
at each fully-connected layer, we apply a Dropout level of $d=0.5$, i.e., during each iteration of training, half of each layer's units do not update their parameters. 
Finally, the Fusing Network output is fed to the last neural network of two units with softmax activation, which yields the probabilities that a particular video is pseudoscientific or not.
We implement our classifier using Keras with Tensorflow as the back-end.

\begin{table}[t!]
\centering
 \setlength{\tabcolsep}{2pt}
 \small
\begin{tabular}{lrrrr}
\toprule
\textbf{Classifier} & \textbf{Accuracy} & \textbf{Precision} & \textbf{Recall} & \textbf{F1 Score}  \\
\midrule
SVM & 0.68 & 0.72 & 0.68 & 0.70 \\
Random Forest & 0.72 & 0.70 & 0.72 & 0.71 \\
BERT-based Classifier & 0.73 & 0.64 & 0.73 & 0.67 \\
\midrule
\textbf{Proposed Classifier} & \textbf{0.76} & \textbf{0.74} & \textbf{0.76} & \textbf{0.74} \\
\midrule
\textbf{Proposed Classifier} & \multirow{2}{*}{\textbf{0.79}} & \multirow{2}{*}{\textbf{0.77}} & \multirow{2}{*}{\textbf{0.79}} & \multirow{2}{*}{\textbf{0.74}} \\
\textbf{(threshold-moving)} & & &  & \\
\bottomrule
\end{tabular}%
\caption{Performance of the evaluated baselines and of the proposed deep learning classifier.}
\label{tab:performance_metrics}
\end{table}

\subsection{Experimental Evaluation}
\label{subsec:experimental_evaluation}
We use ten-fold stratified cross-validation, %
training and testing the classifier for binary classification using all the aforementioned input features.
To deal with data imbalance, we use the Synthetic Minority Over-sampling Technique~\cite{chawla2002smote} and oversample only the training set at each fold.
For stochastic optimization, we use Adam with an initial learning rate of $1\mathrm{e}{-3}$, and $\epsilon=1\mathrm{e}{-8}$.

We then compare the performance of the classifier, in terms of accuracy, precision, recall, and F1 score, using three baselines:
1) a Support Vector Machine (SVM) classifier with parameters $\gamma=0.1$ and $C=10$, 
2) a Random Forest classifier with an entropy criterion and number of minimum samples leaf equal to 2, and
3) a neural network with the same architecture as our classifier that uses a pre-trained BERT model~\cite{turc2019} to learn document-level representations from all the available input features (BERT-based).
For hyper-parameter tuning of baselines (1) and (2), we use the grid search strategy, while for (3), we use the same hyper-parameters as the proposed classifier.
Note that all evaluated models use all available input features.

Table~\ref{tab:performance_metrics} reports the performance of all classifiers.
We observe that our classifier outperforms all baseline models across all performance metrics.
To further reduce false positives and improve the performance of our classifier, we apply a threshold-moving approach, which tunes the threshold used to map probabilities to class labels~\cite{provost2000machine}. 
We use the grid-search technique to find the optimal lower bound probability above which we consider a video pseudoscientific, and find it to be $0.7$.
Using this threshold, we train and re-evaluate the proposed classifier, which yields, respectively, $0.79$, $0.77$, $0.79$, and $0.74$ on the accuracy, precision, recall, and F1 score (see the last row in Table~\ref{tab:performance_metrics}).

\descr{Ablation Study.}
To understand which of the input features contribute the most to the classification of pseudoscientific videos, we perform an ablation study. We systematically remove each of the four input feature types (and their branch in the classifier) and retrain the classifier.
Again, we use ten-fold cross-validation and oversampling to deal with data imbalance and use the classification threshold of $0.7$.
Table~\ref{tab:ablation_study_details} reports the performance metrics for each combination of inputs.
Video tags and transcripts yield the best performance, indicating that they are the most informative input features.
However, using all the available input features yields better performance, which indicates that all four input features are ultimately crucial for the classification task.

\descr{Remarks.} 
Although our classifier outperforms all the baselines, ultimately, its performance ($0.74$ F1-score) reflects the subjective nature of pseudoscientific vs. scientific content classification on YouTube. 
This relates to our crowdsourced annotation's relatively low agreement score, which highlights the difficulty in identifying whether a video is pseudoscientific. It is also evidence of the hurdles in devising models that automatically discover pseudoscientific content.
Nonetheless, we argue that our classifier is only the first step in this direction and can be further improved; overall, it does provide a meaningful signal on whether a video is pseudoscientific (RQ1). 
It can also be used to derive a lower bound of YouTube’s recommendation algorithm's tendency to recommend pseudoscience; we uncover a substantial portion of pseudoscientific videos while also eliminating all false positives with manual review of all the videos classified as pseudoscientific (see Section~\ref{subsec:experiments_design}).

\begin{table}[t!]
\centering
\small
 \setlength{\tabcolsep}{1pt}
\begin{tabular}{lrrrr}
\toprule
\textbf{Input Features} & \textbf{Accuracy} & \textbf{Precision} & \textbf{Recall} & \textbf{F1 Score}  \\
\midrule
Snippet 						& 0.78 & 0.76 & 0.78 & 0.71 \\
Tags 							& 0.78 & 0.77 & 0.78 & 0.72 \\
Transcript 						& 0.78 & 0.74 & 0.78 & 0.71 \\
Comments 						& 0.78 & 0.71 & 0.77 & 0.68 \\
\midrule
Snippet, Tags 					& 0.78 & 0.76 & 0.78 & 0.72 \\
Snippet, Transcript 			& 0.78 & 0.75 & 0.78 & 0.72 \\ 
Snippet, Comments 				& 0.78 & 0.77 & 0.78 & 0.71 \\
Tags, Transcript 				& 0.79 & 0.77 & 0.79 & 0.73 \\ 
Tags, Comments 					& 0.78 & 0.76 & 0.78 & 0.72 \\
Transcript, Comments 			& 0.78 & 0.76 & 0.78 & 0.73 \\ 
\midrule
Snippet, Tags, Transcript 		& 0.78 & 0.75 & 0.78 & 0.72 \\
Snippet, Tags, Comments 		& 0.78 & 0.75 & 0.78 & 0.72 \\ 
Snippet, Transcript, Comments\hspace*{-1cm} 	& 0.78 & 0.76 & 0.78 & 0.73 \\
Tags, Transcript, Comments 		& 0.78 & 0.76 & 0.78 & 0.73 \\ 
\midrule
\textbf{All Features} & \textbf{0.79} & \textbf{0.77} & \textbf{0.79} & \textbf{0.74} \\ 
\bottomrule
\end{tabular}%
\caption{Performance of the proposed classifier (considering the $0.7$ classification threshold) trained with all the possible combinations of the four input feature types.}
\label{tab:ablation_study_details}
\end{table}

\section{Pseudoscientific Content on the YouTube Platform (RQ2)}
\label{sec:analysis}
In this section, we analyze the prominence of pseudoscientific videos on various parts of the platform. %

\subsection{Experimental Design}
\label{subsec:experiments_design}
We focus on three parts of the platform: 1) the homepage; 2) the search results page; and 3) the video recommendations section (recommendations when watching videos). 
Examples of each part of the platform are available from~\cite{anonymouspaperresources}.
We aim to emulate the logged-in and non-logged-in users' behavior with varying interests and measure how the watch history affects pseudoscientific content recommendation.
To do so, we create three different Google accounts, each one with a different watch history, while all the other account information is the same to avoid confounding effects caused by profile differences.
Additionally, we perform experiments on a browser without a Google account to emulate not logged-in users. 
Moreover, we perform experiments using the YouTube Data API (when the API provides the required functionality) to investigate the differences between YouTube as an application and the API.

\begin{algorithm}[t!]
\centering
\small
 \setlength{\tabcolsep}{1pt}
\caption{Minimum number of videos needed to build the watch history of a user profile.}
\label{alg:findminimunwvideosalgorithm}
\begin{algorithmic}[1]
	\State Let $\mathbf{S}$ be a set of 100 randomly selected COVID\-19 pseudoscientific videos
	\State Let $\mathbf{V}_{ref}$ be a randomly selected COVID\-19 pseudoscientific video
	\State Let $\mathbf{V}_{refRec}$ be the top $10$ recommendations of $\mathbf{V}_{ref}$
	\State $\mathbf{RH}_{recs} \gets \{\mathbf{V}_{refRec}$\}
	\State $\mathbf{S}_{threshold} \gets 1.0$
	\State $\mathbf{W} \gets 0$ \Comment{Number of videos watched}
	\For{\textbf{each} video $\mathbf{V}$ in $\mathbf{S}$}
		\State Watch video $\mathbf{V}$
		\State $W\gets W + 1$
		\State Get the top 10 recommendations $\mathbf{R}$ of $\mathbf{V}_{ref}$
		\State Calculate the Overlap Coefficient $\mathbf{O}_{coef}$ between \\
		\hskip\algorithmicindent $\mathbf{R}$ and $\mathbf{RH}_{recs}$
		\If {$\mathbf{O}_{coef} \geq \mathbf{S}_{threshold}$}
			\State \textbf{return} $\mathbf{W}$
		\Else
			\State Add $\mathbf{R}$ to the set of recommendations $\mathbf{RH}_{recs}$ \\
			\hskip\algorithmicindent\hskip\algorithmicindent retrieved in the previous iterations
		\EndIf
	\EndFor
\end{algorithmic} %
\end{algorithm}

\descr{User Profile Creation.}
According to Hussein et al.~\cite{hussein2020measuring}, once a user forms a watch history, user profile attributes (i.e., demographics) affect future video recommendations.
Hence, since we are only interested in the watch history, each of the three accounts has the same profile: 30 years old and female. 
To decrease the likelihood of Google automatically detecting our user profiles, we carefully crafted each one assigning them a unique name and surname 
and performed standard phone verification.
None of the created profiles were banned or flagged by Google during or after our experiments.

\descr{Watch History.} 
We build the watch history of each profile, aiming to create the following three profiles: 
1)~a user interested in legitimate science videos (``Science Profile'');
2)~a user interested in pseudoscientific content (``Pseudoscience Profile''); and
3)~a user interested in both science and pseudoscience videos (``Science/Pseudoscience Profile'').
To find the minimum number of videos a profile needs to watch before YouTube learns the user's interests and starts generating more personalized recommendations, 
we use a newly created Google account with no watch history, and we devise and execute the following algorithm (see Algorithm~\ref{alg:findminimunwvideosalgorithm}).
First, we randomly select a video, which we refer to as the ``reference'' one, from the COVID-19 pseudoscientific videos of our ground-truth dataset, and we collect its top 10 recommended videos.
Next, we create a list of 100 randomly selected COVID-19 pseudoscientific videos, excluding videos exceeding five minutes in duration, and we repeat the following process iteratively:

\begin{enumerate}
\item We start by watching a video from the list of the randomly selected pseudoscientific videos;

\item We visit the reference video, and we collect the top 10 recommendations, store them, and compare them using the Overlap Coefficient with all the recommendations of the reference video collected in the previous iterations;

\item If all the recommended videos of the reference video at the current iteration have also been recommended in the previous iterations (Overlap Coefficient = 1.0), we stop our experiment. Otherwise, we increase the number of videos watched and proceed to the next iteration.
\end{enumerate}

\begin{figure}[t!]
\centering
\includegraphics[width=0.95\columnwidth]{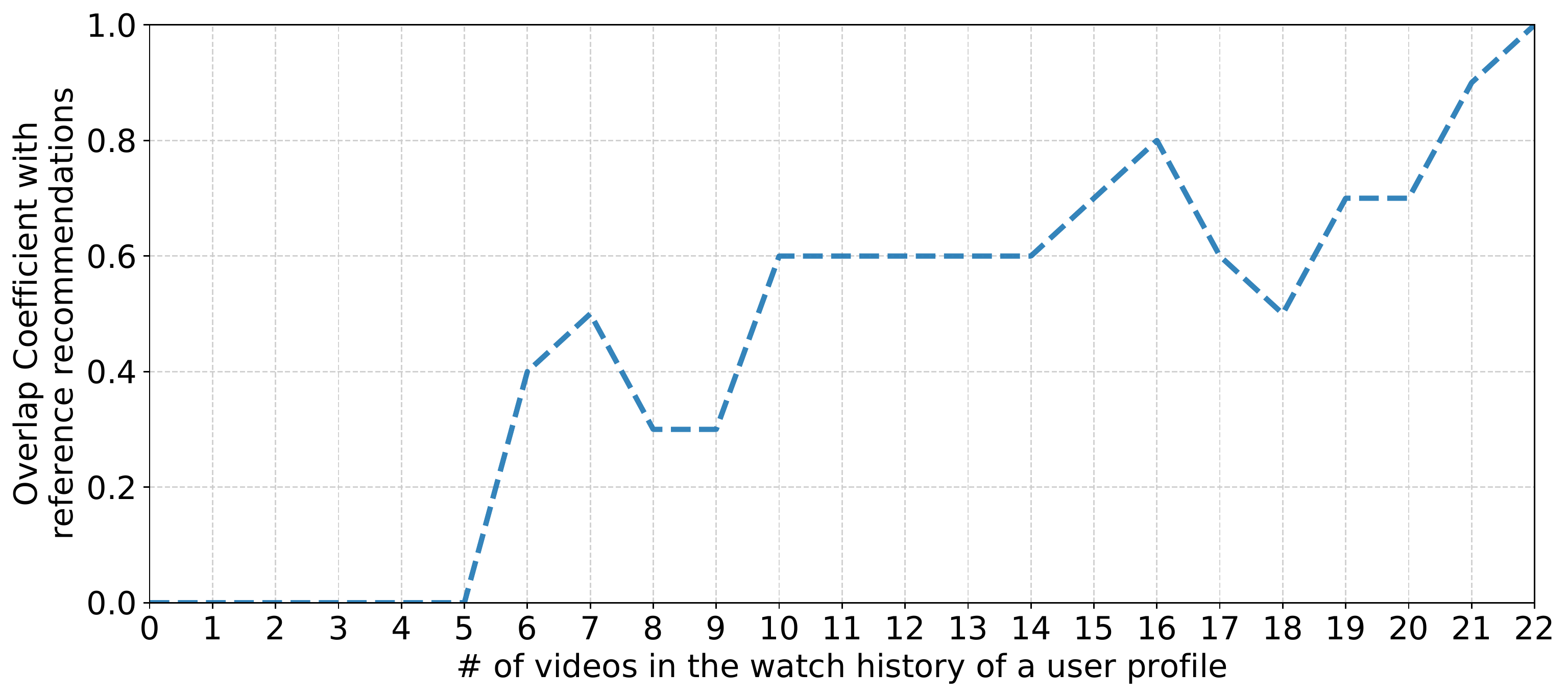}
\caption{Overlap Coefficient with the reference recommendations of previous iterations of the devised algorithm as a function of the number of videos in the user's watch history.}
\label{fig:similarity_build_watch_history}
\end{figure}

Using this algorithm, we find that the minimum amount of videos required to be watched by a user for YouTube to start generating more personalized recommendations is 22.
Figure~\ref{fig:similarity_build_watch_history} depicts the overlap coefficient between the recommendations at each iteration and the reference recommendations of previous iterations as a function of the number of videos in the user's watch history.
However, to create more representative watch histories and get even more personalized recommendations, we increase this number to 100. %
Finally, we select the most popular science and pseudoscience videos from the ground-truth dataset, based on the number of views, likes, comments, etc., and use them to personalize the three Google accounts' profiles.
Since it is not clear how YouTube measures the satisfaction score on videos and how watch time affects this score, during profile training, we always watch the same proportion of the video ($50\%$ of the total duration).

\revisionfinal{
Unlike Hussein et al.~\cite{hussein2020measuring}, we decide not to take into account the rankings of the videos for our calculations.
When it comes to the amount of scientific/pseudoscientific content being presented to the user it is unclear how the users' watching selections are affected by the ranking.
We believe that our results and reported percentages are still highly indicative.
}
\revcomment{(Comment: R.3.2)}

\descr{Controlling for noise.}
Some differences in search results and recommendations are likely due to factors other than the user's watch history and personalization in general.
To reduce the possibility of this noise affecting our results, we take the following steps: 
1)~We execute, in parallel, experiments with identical search queries for all accounts to avoid updates to search results over time for specific search queries;
2)~All requests to YouTube are sent from the same geographic location (through the same US-based Proxy Server) to avoid location-based differentiation;
3)~We perform all experiments using the same browser user-agent and operating system;
4)~To avoid the carry-over effect (previous search and watch activity affecting subsequent searches and recommendations), at each repetition of our experiments, we use the ``Delete Watch and Search History'' function %
to erase the activity of the user on YouTube from the date after we built the user profiles; and
5)~Similarly to the profiles' watch history creation, we always watch the same proportion of the video ($50\%$ of the total duration).
\descr{Implementation.} The experiments are written as custom scripts using Selenium in Python 3.7.
For each Google account, we create a separate Selenium instance for which we set a custom data directory, thus being able to perform manual actions on the browser before starting our experiments, e.g., performing Google authentication, installing AdBlock Plus
to prevent advertisements within YouTube videos from interfering with our emulations, etc.
Finally, for all our experiments, we use Chromedriver 83.0.4 that runs in headless mode and stores all received cookies.

\descr{Video Annotation.} 
Here, we describe how we use our classifier in our experiments.
In particular, we initially use our classifier to annotate all the videos encountered in our experiments and identify videos that are more likely to be pseudoscientific.
Then, the first author of this paper manually inspects all the videos classified as pseudoscientific to confirm that they are indeed pseudoscientific. 
Following this approach, we eliminate all the false positives.

\begin{figure}[t!]
\centering
\includegraphics[width=0.85\columnwidth]{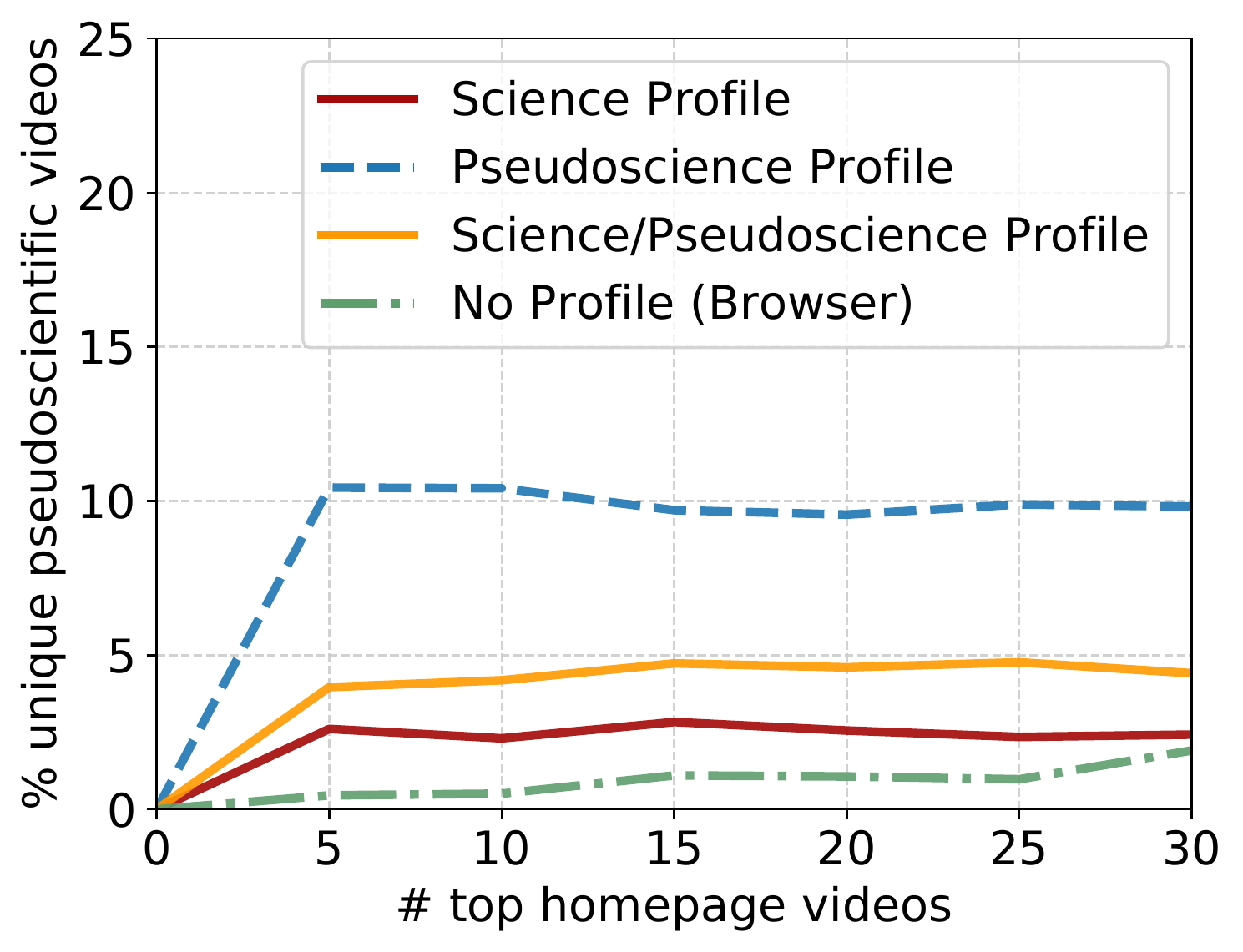}
\caption{Percentage of pseudoscience videos found in the homepage of each user profile.}
\label{fig:homepage_experiment_plot}
\end{figure}

\begin{figure*}[t!]
\centering
\includegraphics[width=0.75\linewidth]{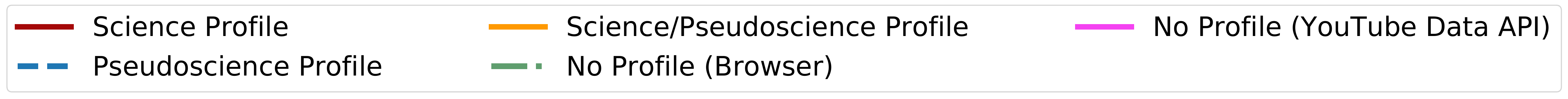}\\
\includegraphics[width=0.23\linewidth]{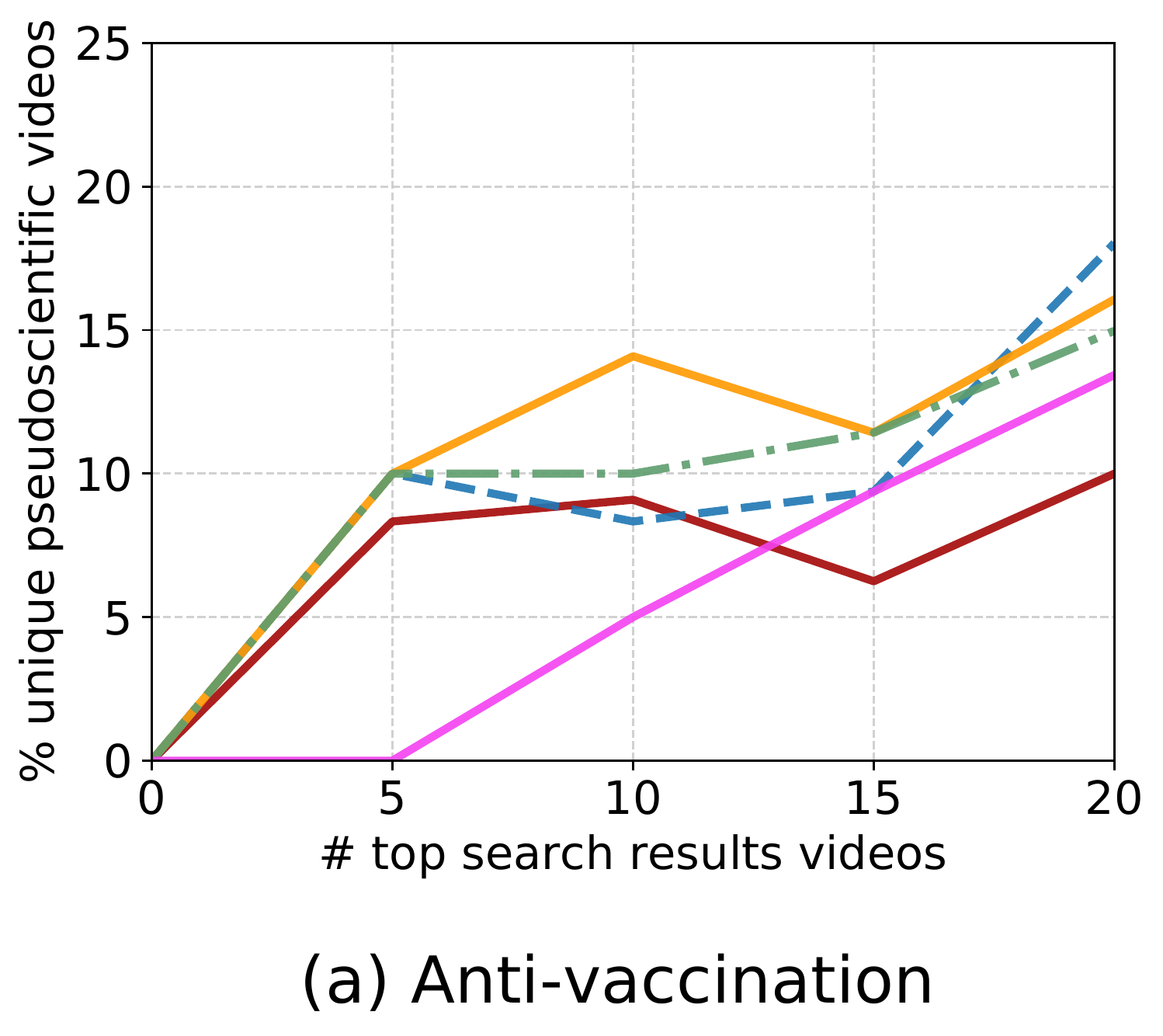}
\includegraphics[width=0.23\linewidth]{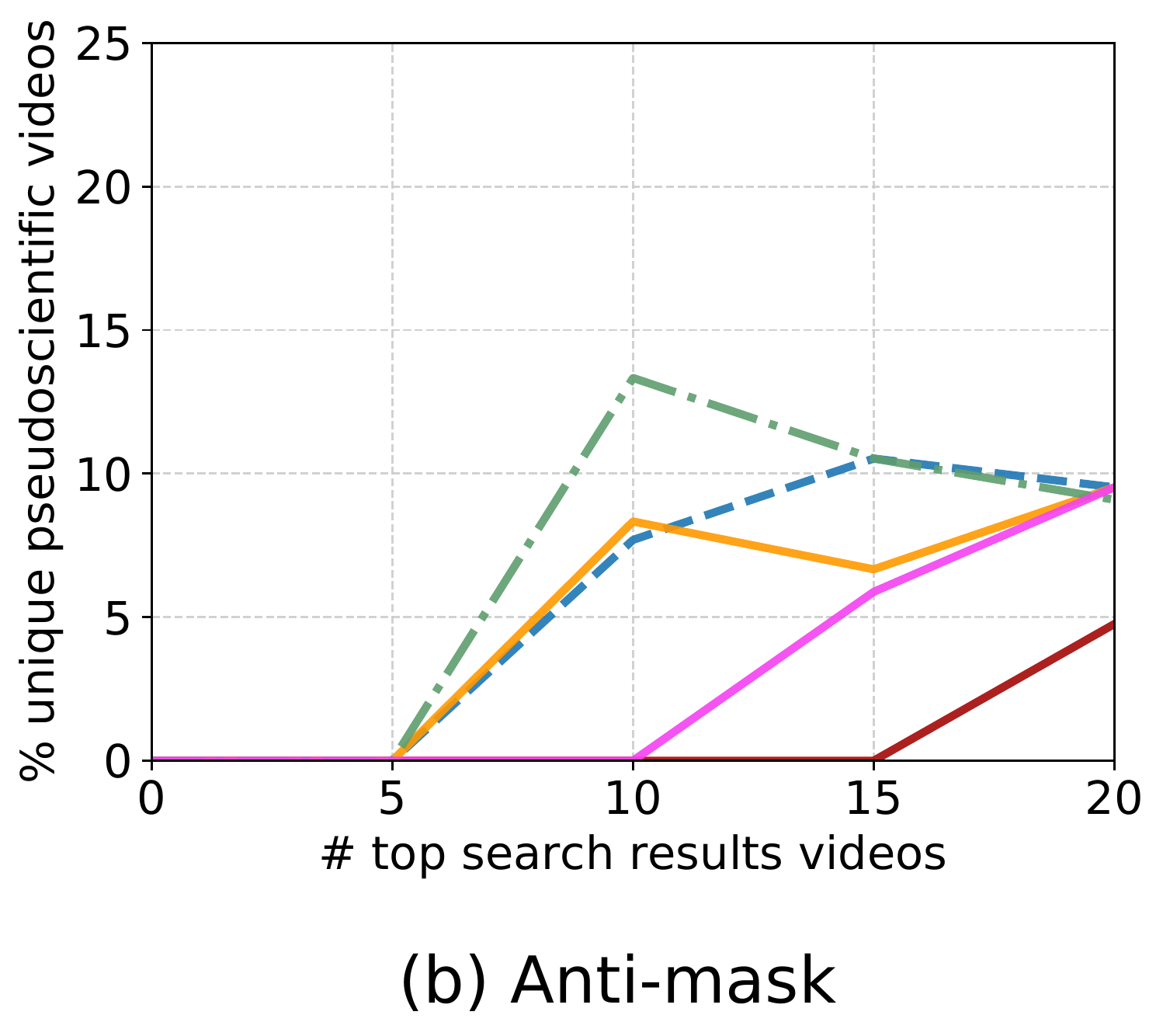}
\includegraphics[width=0.23\linewidth]{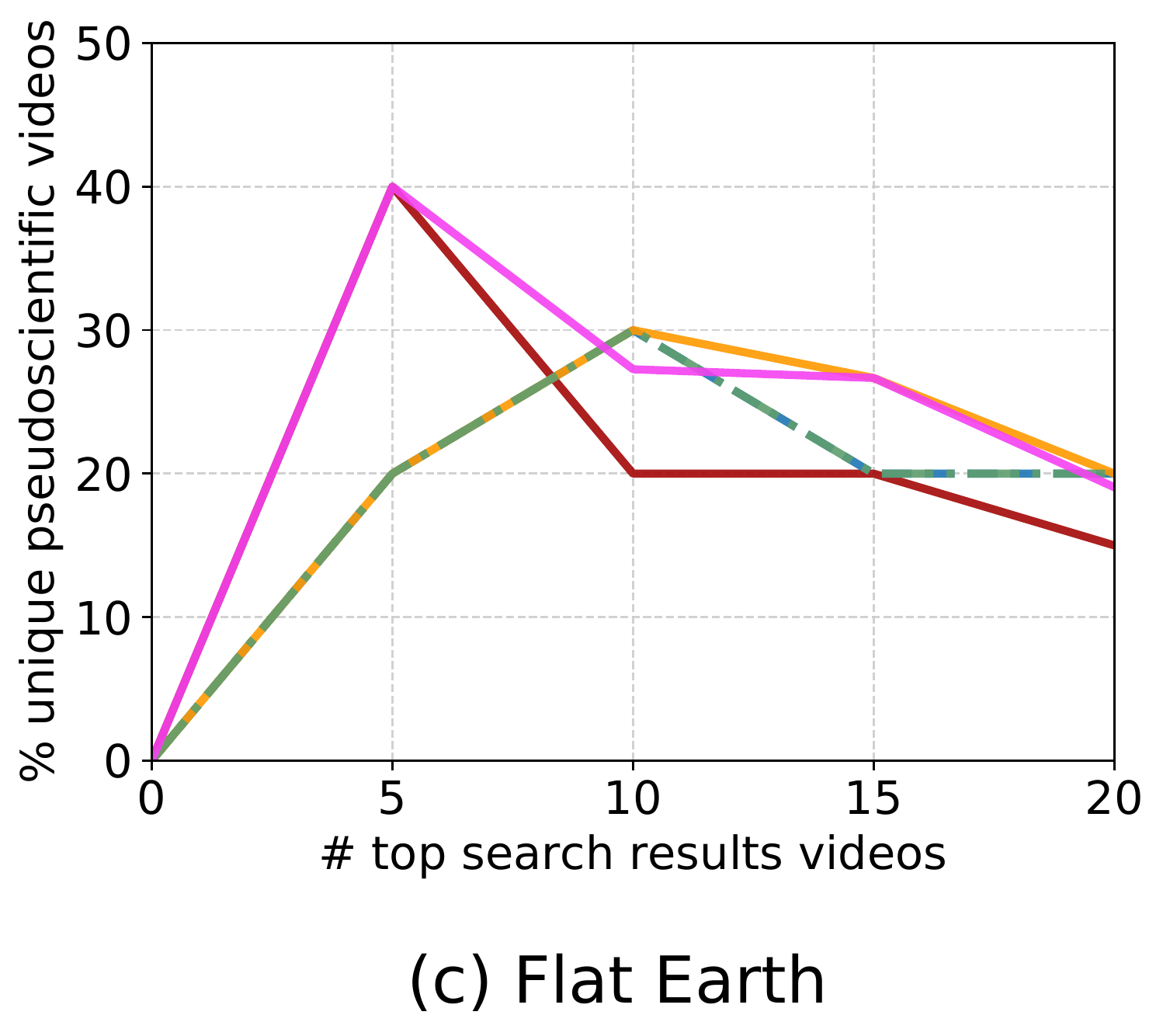}
\caption{Percentage of unique pseudoscience videos found in the search results of each user profile.}
\label{fig:youtube_search_experiment_plot_all_terms}
\end{figure*}

\subsection{Pseudoscientific Content on Homepage, Search Results, and Video Recommendations}
\label{subsec:experiments_details}
\descr{Homepage.}
We begin by assessing the magnitude of the pseudoscientific content problem on the YouTube homepage.
To do so, we use each one of the three user profiles (Science, Pseudoscience, and Science/Pseudoscience), as well as another user with no account (No Profile) that emulates the behavior of not logged-in users. 
We then visit each profile's homepage to collect and classify the top 30 videos as ranked by YouTube.
Note that we cannot perform this experiment using the YouTube Data API since it does not support this functionality. %
We repeat the same experiment 50 times with a waiting time of 10 minutes between each repetition because YouTube shows different videos on the homepage each time a user visits YouTube.
We perform this experiment during December, 2020.

Figure~\ref{fig:homepage_experiment_plot} shows the percentage of unique pseudoscientific videos on the homepage of each user profile.
We find that $2.4\%$, $9.8\%$, $4.4\%$, and $1.9\%$ of all the unique videos found in the top 30 videos of the homepage of the  Science, Pseudoscience, Science/Pseudoscience, and the No profile (browser) users, respectively, are pseudoscientific.
Overall, the Pseudoscience and the Science/Pseudoscience profile receive a higher percentage of pseudoscientific content.%
We also verify the significance of the difference in the amount of pseudoscientific content in the homepage of the Pseudoscience and the Science/Pseudoscience profiles compared to the one of the No profile (browser) using the Fisher's Exact test ($p<0.05$).%
We obtain similarly high significance ($p<0.05$) when we compare the Pseudoscience and Science/Pseudoscience profiles with the Science profile.
This indicates that the users' watch history substantially affects the number of pseudoscientific recommendations on their homepage. Nevertheless, users who are not interested in this type of content (i.e., science profile) still receive a non-negligible amount of pseudoscientific content.
We also observe that as the number of videos on the user's homepage increases (e.g., when a user scrolls down), the pseudoscientific videos' percentage remains approximately identical.

\begin{figure*}[t!]
\centering
\includegraphics[width=0.75\linewidth]{figures/random_walks_legend.pdf}\\
\includegraphics[width=0.23\linewidth]{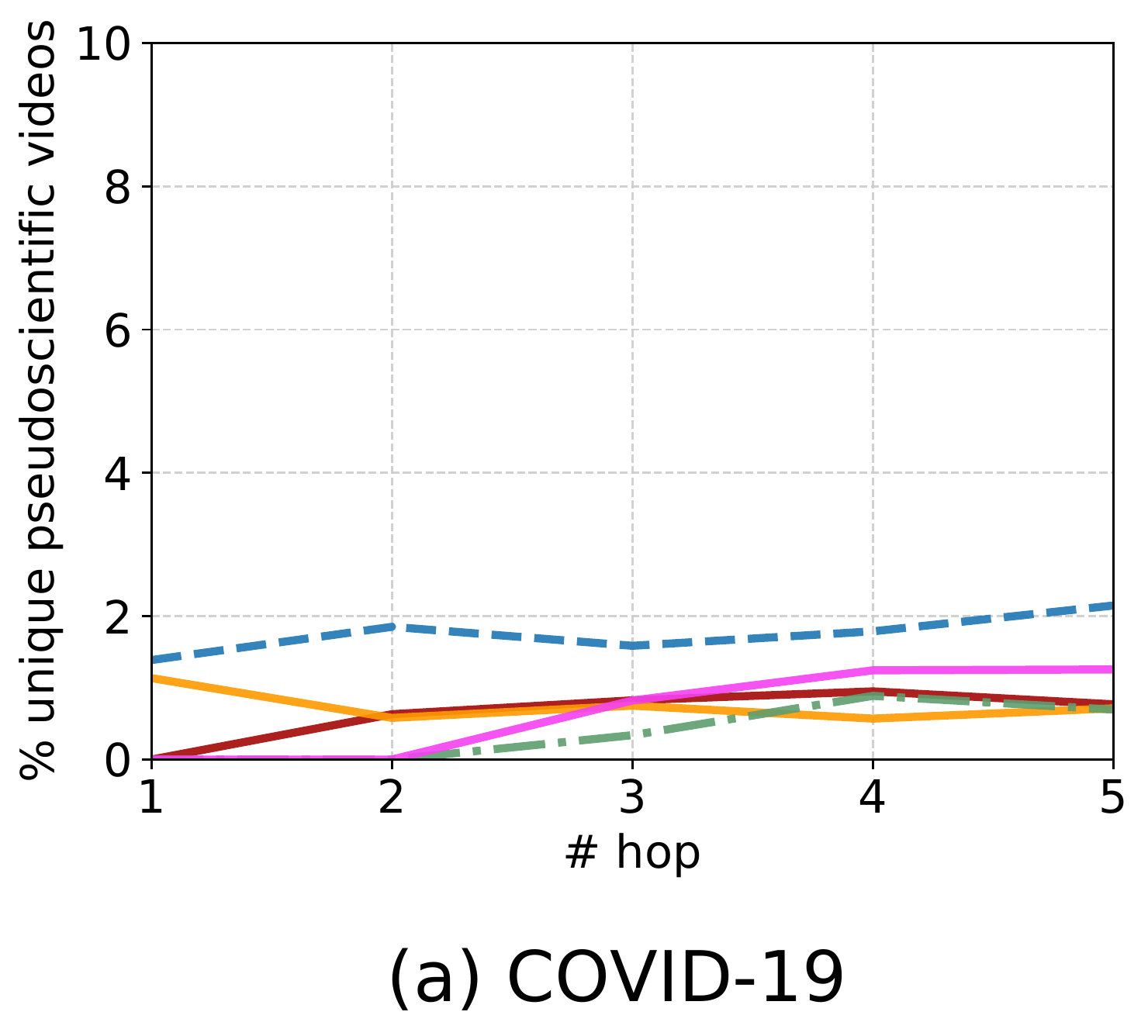}\label{fig:random_walks_covid}
\includegraphics[width=0.23\linewidth]{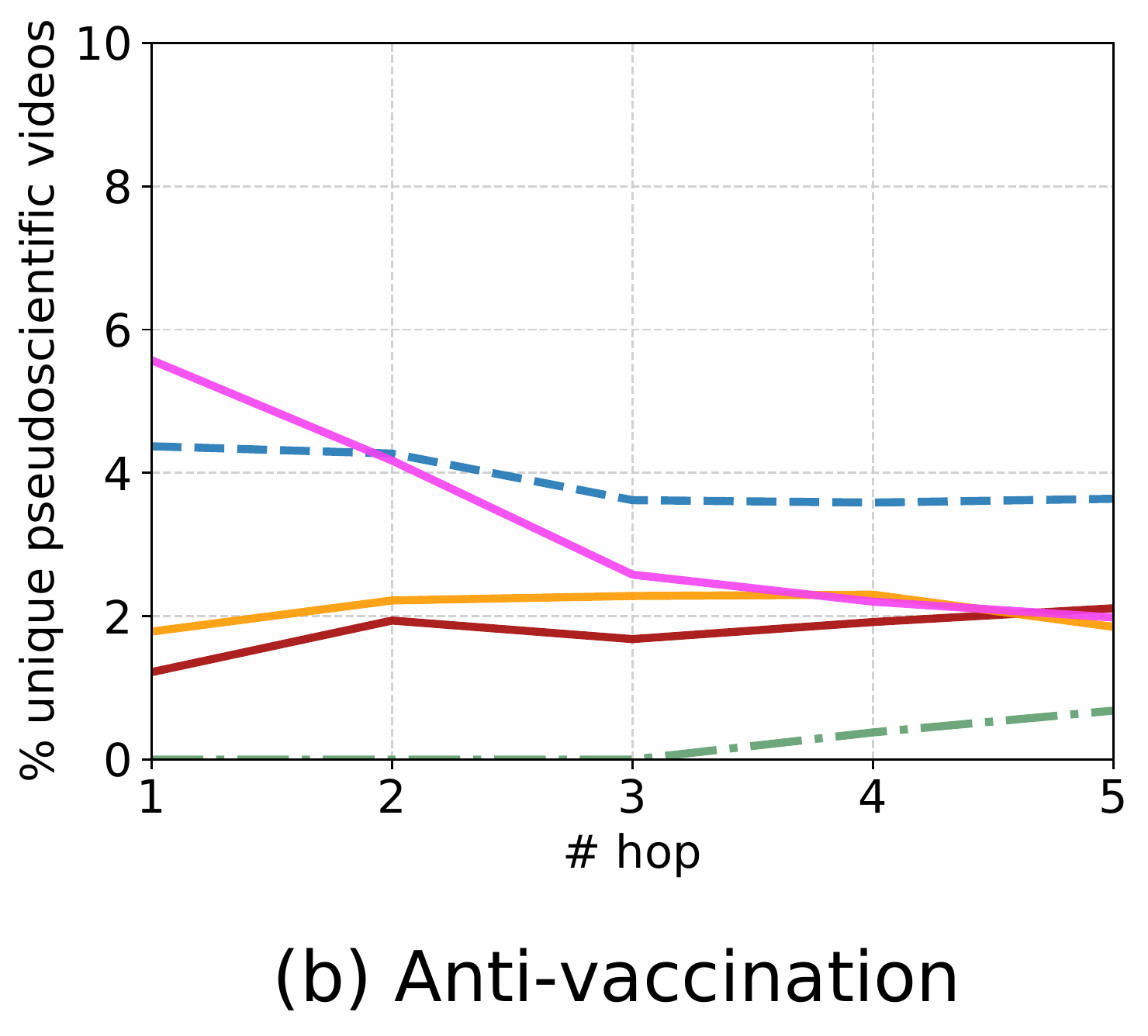}
\includegraphics[width=0.23\linewidth]{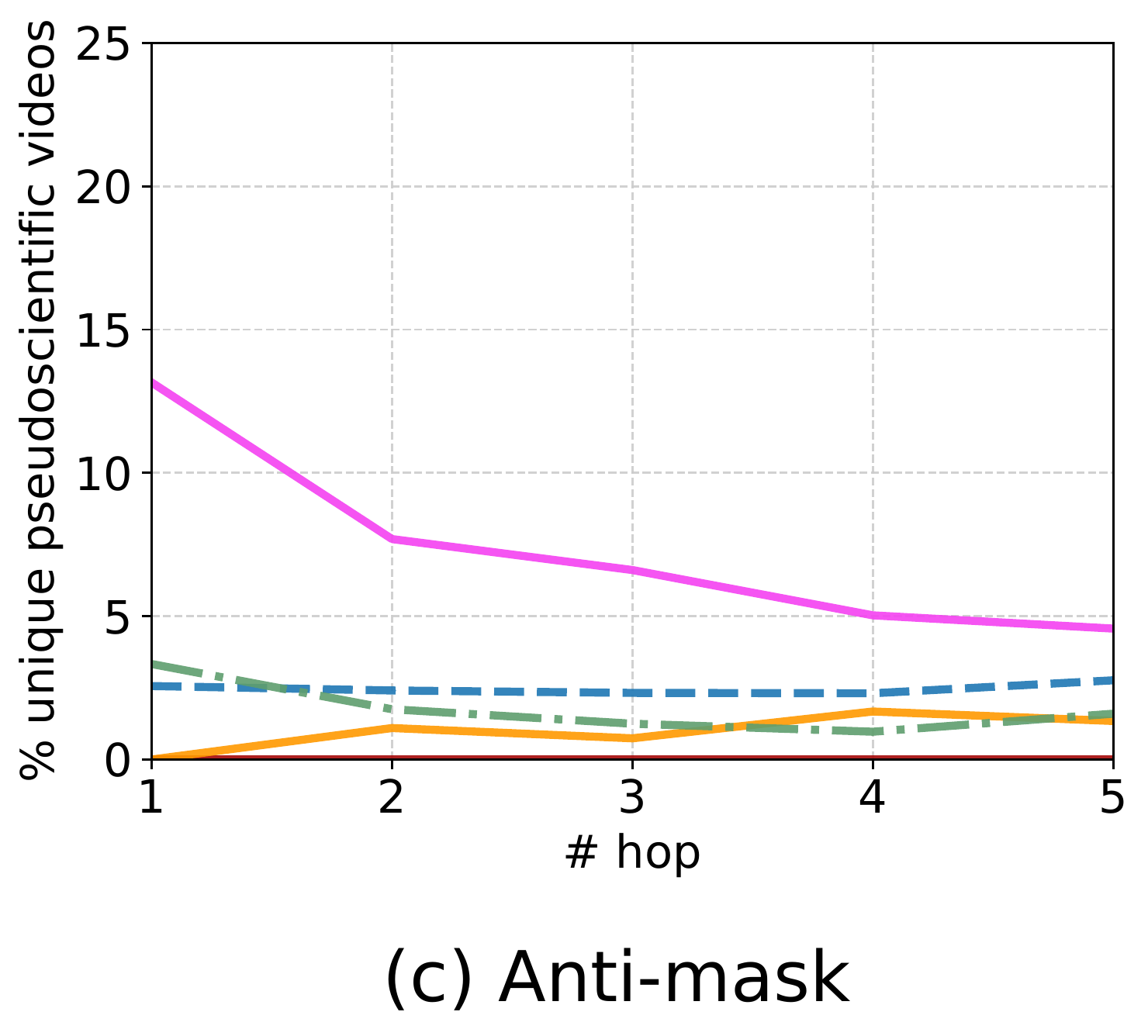}
\includegraphics[width=0.23\linewidth]{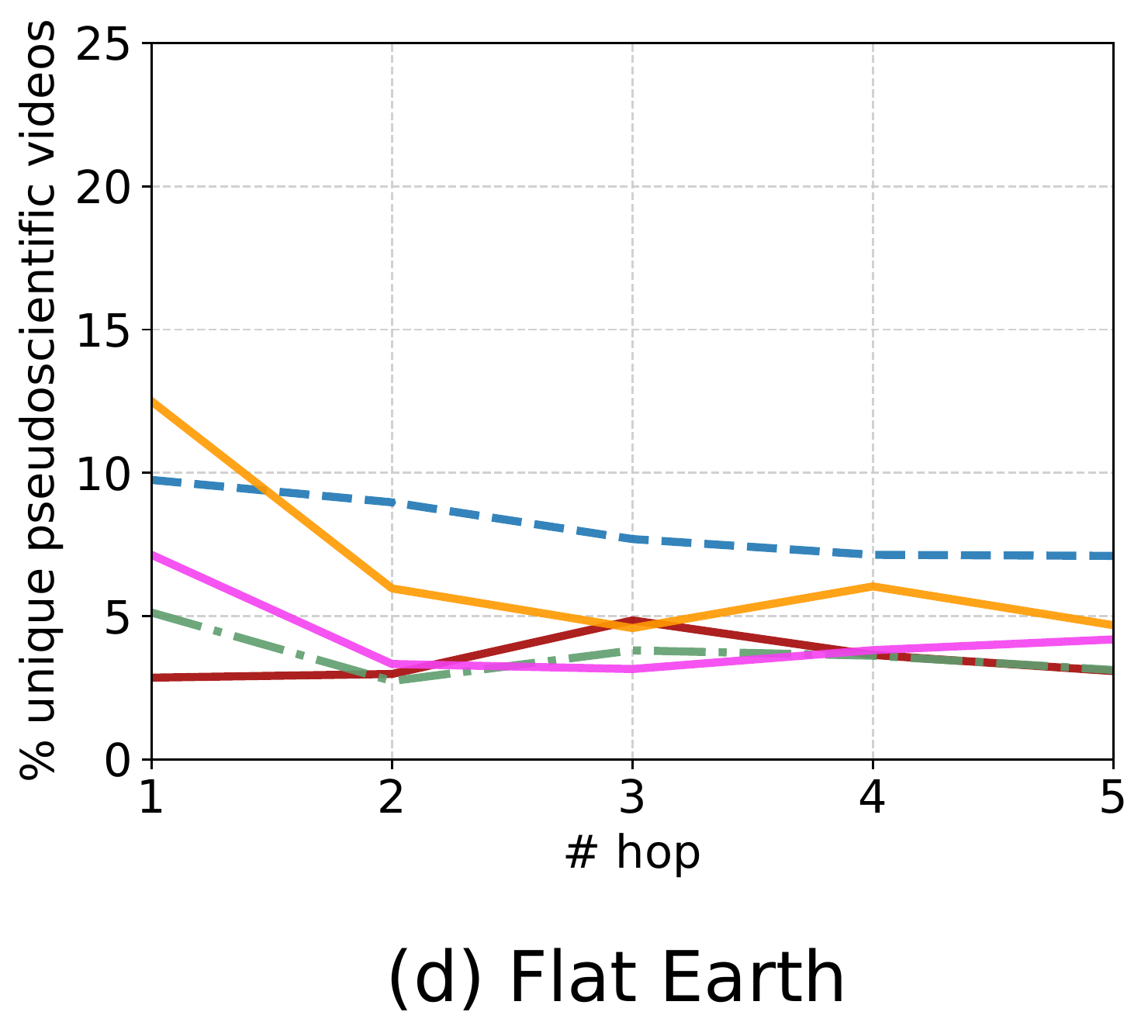}
\caption{Percentage of unique pseudoscientific videos that the random walker encounters at hop $k$ per user profile (December 2020).}
\label{fig:random_walks_pseudoscience_found}
\end{figure*}

\descr{Search Results.}
Next, we focus on quantifying the prevalence of pseudoscientific content when users search for videos on YouTube.
For this experiment, we perform search queries on YouTube using the four pseudoscientific topics in our ground-truth dataset.
For topics with two search queries (i.e., COVID-19), we perform the experiment twice and average their results.
We retrieve the top 20 videos for each search query and use our classifier to classify each video in the result set.
We repeat this experiment 50 times for each pseudoscientific topic using all three user profiles and two non-logged-in users with no profile (one using a browser and another using YouTube's Data API).
Recall that we delete the user's watch history at each experiment repetition and between those performed with different search queries to ensure that future search results are not affected by previous activity other than our controlled watch history.
We perform this experiment in December, 2020.

Overall, we find a large variation in the results across pseudoscientific topics (see Fig.~\ref{fig:youtube_search_experiment_plot_all_terms}).
For more traditional pseudoscientific topics like Flat Earth, YouTube search returns even more pseudoscientific content. %
In particular, when searching for Flat earth, the Science profile, Pseudoscience profile, Science/Pseudoscience profile, no profile (browser), and the Data API encounter, respectively, $5.0\%$, $2.0\%$, $3.9\%$, $5.0\%$, and $5.6\%$ more unique pseudoscientific content than when searching for Anti-vaccination.
In fact, Anti-vaccination is the topic with the second-highest amount of pseudoscientific content across all profiles.
For topics like COVID-19, all the recommended videos are \emph{not} pseudoscientific, suggesting that YouTube's recommendation algorithm does a better job in recommending less harmful videos---at least for COVID-19.
This also signifies that YouTube has made substantial efforts to tackle COVID-related misinformation~\cite{youtubecovid2020tackle}, establishing an official, dedicated policy for that~\cite{youtubecovid2020policy}.
However, this is not the case for other controversial and timely pseudoscientific topics like Anti-vaccination or Anti-mask.
An explanation of the differences observed between COVID-19 and Anti-mask lies in that COVID-19 has a longer timeline than the masks-related problem.
The Anti-mask movement gained attraction after a few months from the emergence of the COVID-19 pandemic and YouTube might need some more time to develop effective moderation strategies to tackle misinformation surrounding the use of masks.
Nevertheless, YouTube has recently announced that they will also attempt to target COVID-19 vaccine misinformation~\cite{youtubevaccine2020tackle}.

For Anti-vaccination, Anti-mask, and Flat earth searches, YouTube outputs more pseudoscientific content to the Pseudoscience and Science/Pseudoscience profiles than to the Science one.
Specifically, the amount of unique pseudoscientific videos in the top 20 search results of the Pseudoscience profile is, respectively, $18.0\%$, $9.5\%$, and $20.0\%$ for Anti-vaccination, Anti-mask, and Flat Earth.
For the Science/Pseudoscience profile, it is $16.1\%$, $9.5\%$, and $20.0\%$, while for the Science one is $10.0\%$, $4.8\%$, and $18.0\%$.

Furthermore, when taking into account the ranking of the search results, as the number of search results increases for Anti-vaccination and Anti-mask so does the percentage of unique pseudoscientific videos, which might indicate that YouTube does a good job in ranking content with higher quality on top for this topics.
On the other hand, for Flat Earth more of the pseudoscientific content is observed in the top five search results.
\descr{Video Recommendations.}
\label{subsec:random_walks_experiment}
Last but not least, we set out to assess YouTube's recommendation algorithm's pseudoscience problem by performing controlled, live random walks on the recommendation graph while again measuring the effect of a user's watch history. 
This allows us to emulate the behavior of users with varying interests who search the platform for a video and subsequently watch several videos according to recommendations.
Note that videos are nodes in YouTube's recommendation graph, and video recommendations are directed edges connecting a video to its recommended videos.
For example, a YouTube video page can be seen as a snapshot of YouTube's recommendation graph showing a single node (video) and all the directed edges to all its recommended videos in the graph.

For our experiments, we use the four pseudoscientific topics considered for the creation of our ground-truth dataset. We initially perform a search query on YouTube and randomly select one video from the top 20 search results for each topic. 
We then watch the selected video, obtain its top ten recommended videos, and randomly select one. 
Again, we watch that selected video and randomly choose one of its top 10 recommendations.
This emulates the behavior of a user who watches videos based on recommendations, selecting the next video randomly from among the top 10 recommendations until he reaches five hops (i.e., six total videos viewed), thus ending a single live random walk.
We repeat this process for 50 random walks for each search query related to each topic while automatically classifying each video we visit. %
For topics with two search queries (i.e., COVID-19), we perform the experiment twice and average their results.
We also ensure that the same video is not selected twice within the same random walk and that all random walks of a user profile performed for the same topic are unique.
We perform this experiment with all user profiles and the API during December, 2020.
Note that the recommendations collected using the API differ from the recommendations collected from a browser.
In fact, the API allows us to collect the ``related'' videos of a given video, which are recommendations provided by YouTube's recommendation algorithm based on video item-to-item similarity, as well as general user engagement and satisfaction metrics.
Second, the API does not provide a functionality to watch YouTube videos. 

\begin{table}[t!]
\centering
\small
\resizebox{1\columnwidth}{!}{%
 \setlength{\tabcolsep}{2pt}
\begin{tabular}{llrrrrrrr}
\toprule
& & \multicolumn{3}{c}{\bf Profile} & \multicolumn{2}{c}{\bf No Profile} \\
& & \textbf{Sci} & \textbf{Pseudo} & \textbf{Sci/Pseudo} & \textbf{Browser} & \textbf{API}  \\
\toprule
\textbf{Home} & - & $2.4\%$ & $9.8\%$ & $4.4\%$ & $1.9\%$ & - \\
{\bf (Top 30)}\\
\midrule
\multirow{5}{*}{\begin{tabular}[c]{@{}l@{}}\textbf{Search}\\ \textbf{(Top 20)}\end{tabular}} 
& COVID-19 & 0.0\% & 0.0\% & 0.0\% & 0.0\% & 0.0\% \\
 & Anti-vacc & 10.0\% & 18.0\% & 16.1\% & 15.0\% & 13.4\% \\
 & Anti-mask & 4.8\% & 9.5\% & 9.5\% & 9.1\% & 10.0\% \\
 & Flat Earth & 15.0\% & 20.0\% & 20.0\% & 20.0\% & 19.0\% \\
 & All Studied & 6.6\% & 10.9\% & 10.3\% & 9.8\% & 9.2\% \\
 & Topics & & & & \\
\midrule
\multirow{5}{*}{\begin{tabular}[c]{@{}l@{}}\textbf{Video}\\ \textbf{Recs}\end{tabular}}
 & COVID-19 & 0.8\% & 2.1\% & 0.7\% & 0.7\% & 1.3\% \\
 & Anti-vacc & 2.1\% & 3.6\% & 1.9\% & 0.7\% & 2.0\% \\
 & Anti-mask & 0.0\% & 2.8\% & 1.3\% & 1.6\% & 4.6\% \\
 & Flat Earth & 3.1\% & 7.1\% & 4.7\% & 3.1\% & 4.2\% \\
 & All Studied & 1.5\% & 3.6\% & 1.9\% & 1.2\% & 2.5\% \\
 & Topics & & & & \\
\bottomrule
\end{tabular}%
}
\caption{Percentage of unique pseudoscientific videos encountered by each user profile in the three main parts of YouTube.}
\label{tab:experiments_pseudoscience_videos_found_details}
\end{table}

\begin{figure*}[t!]
\centering
\includegraphics[width=0.75\linewidth]{figures/random_walks_legend.pdf}\\
\includegraphics[width=0.23\linewidth]{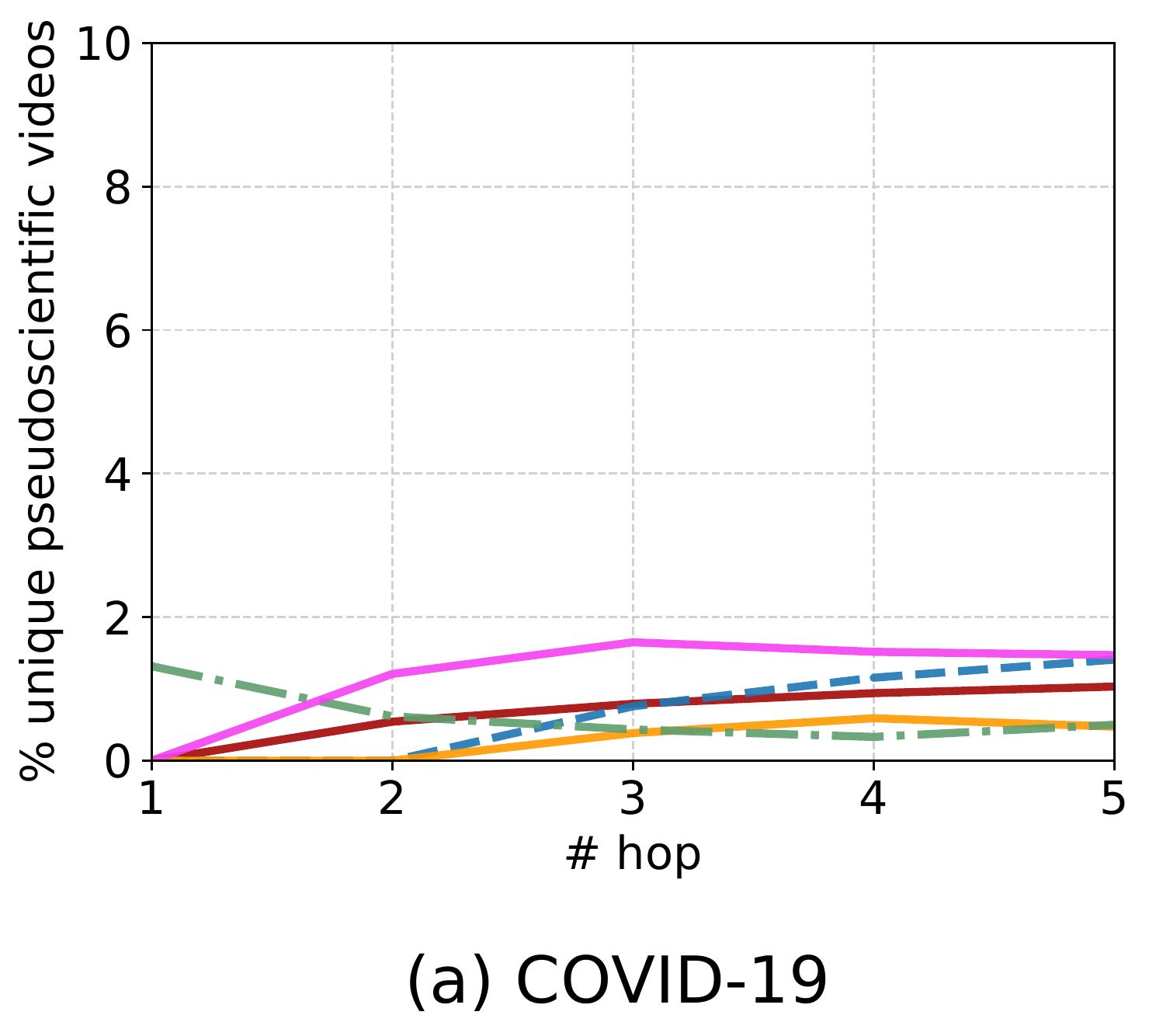}
\includegraphics[width=0.23\linewidth]{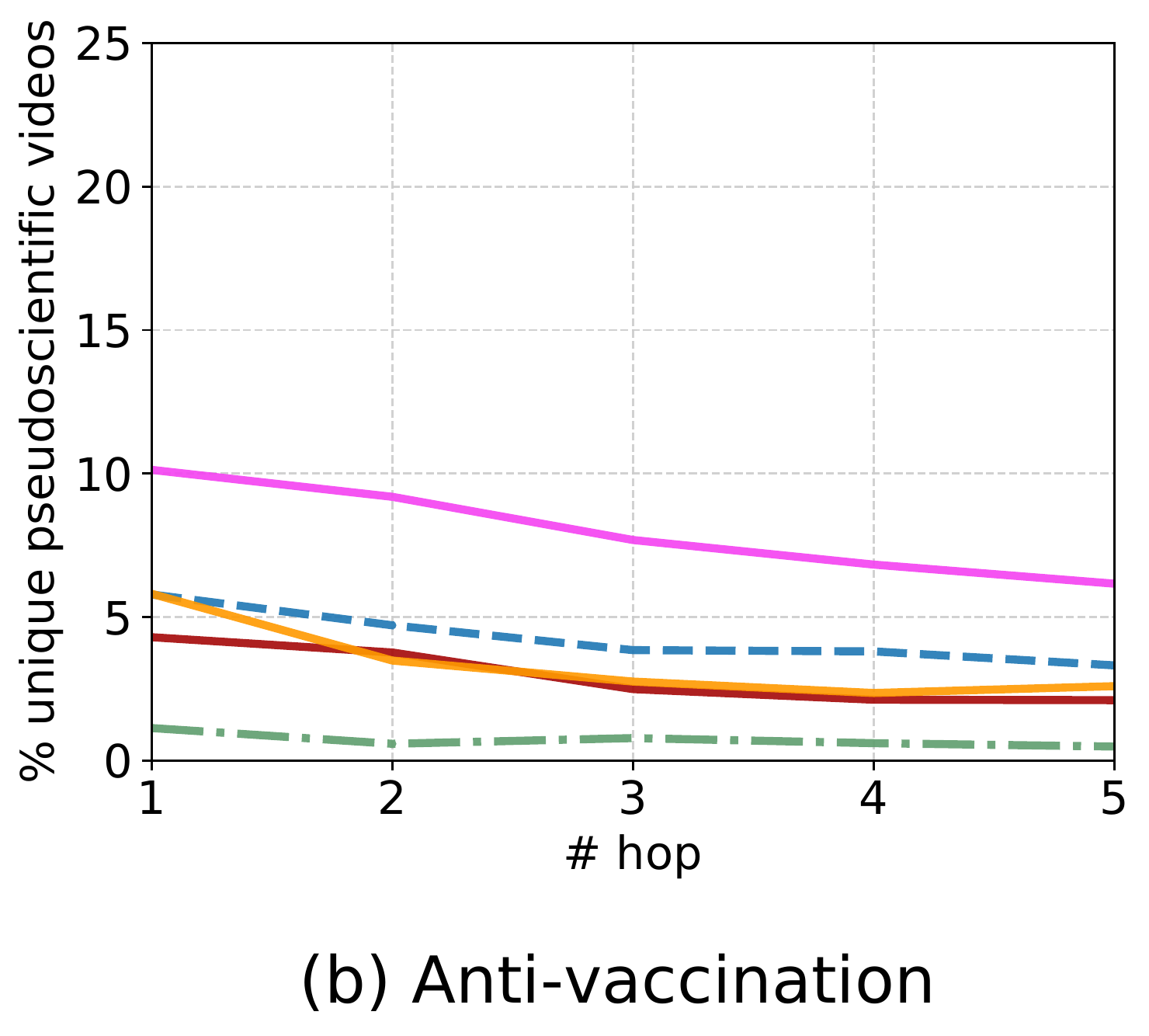}
\includegraphics[width=0.23\linewidth]{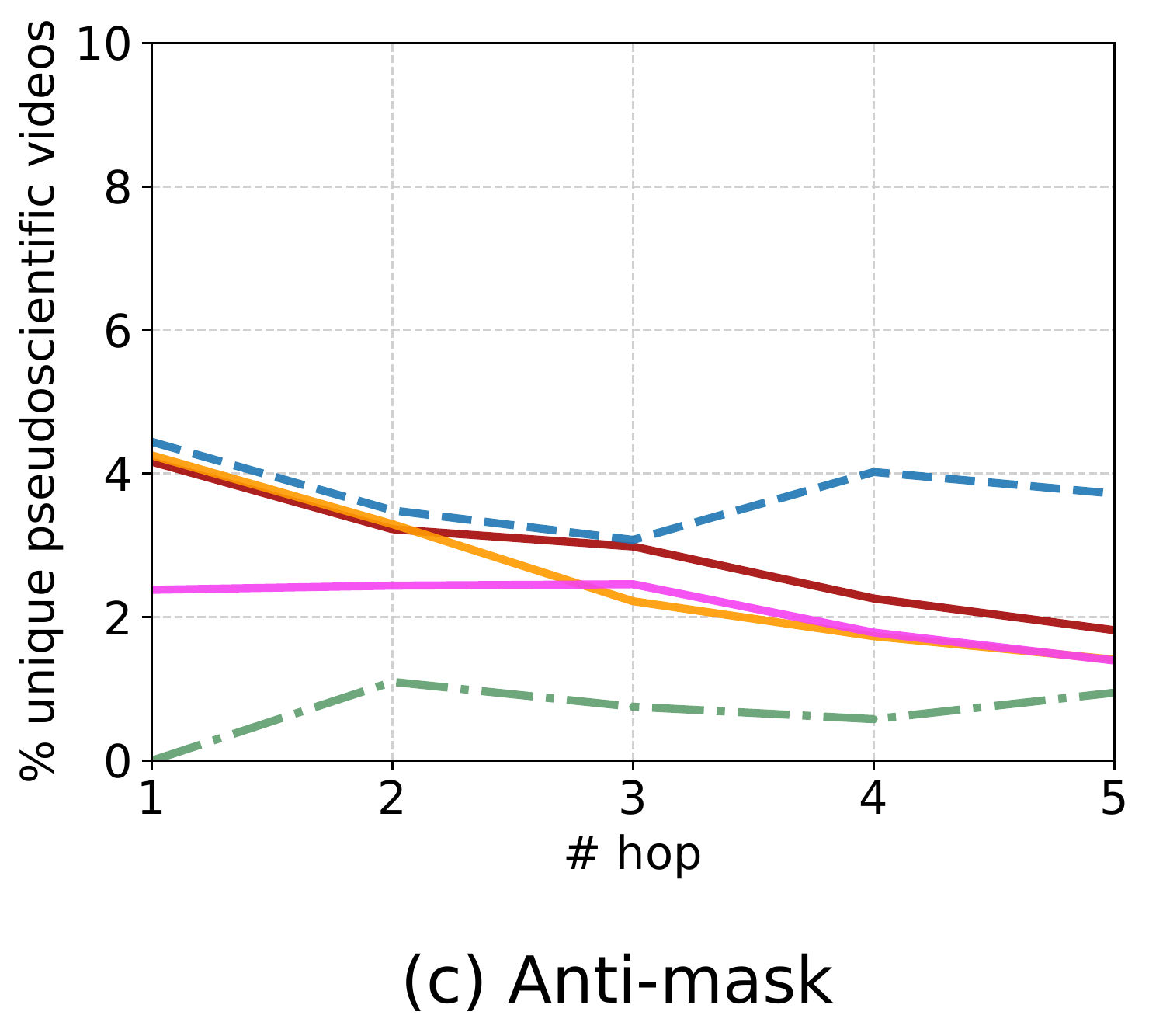}
\includegraphics[width=0.23\linewidth]{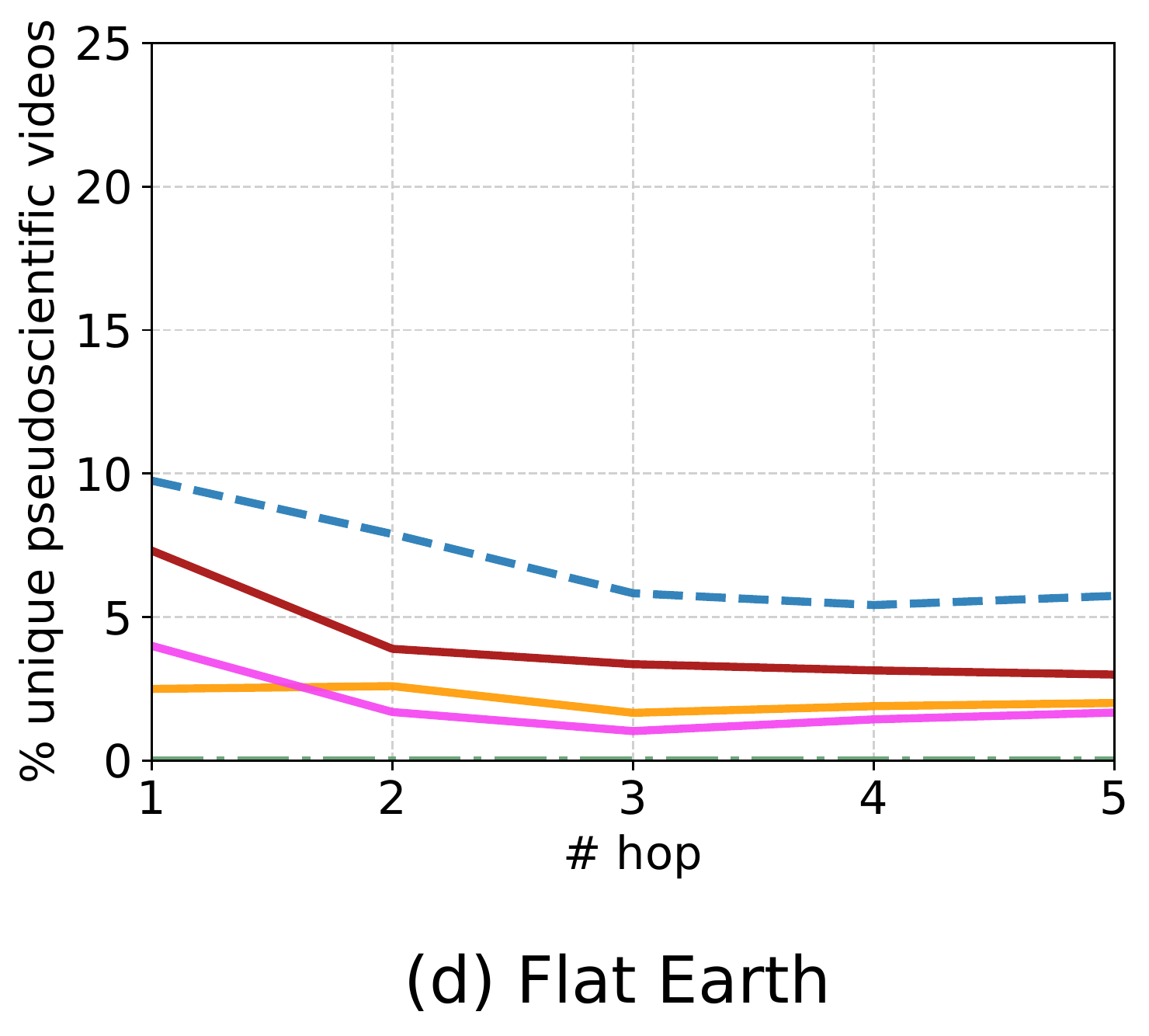}
\caption{Percentage of unique pseudoscientific videos that the random walker encounters at hop $k$ per user profile (April-May 2021).}
\label{fig:random_walks_pseudoscience_found_second}
\end{figure*}

For each user profile's random walks, we calculate the percentage of pseudoscientific videos encountered over all unique videos that the random walker visits up to the k-th hop. 
Note that we have already assessed the amount of pseudoscientific content in the search results. 
Hence, in this experiment, we focus on video recommendations and do not consider, in our calculations, the initial video of each random walk selected from the search results.

Figure~\ref{fig:random_walks_pseudoscience_found} plots this percentage per hop for each of the pseudoscientific topics explored.
Looking at the percentage of pseudoscientific videos encountered by each user profile in all the random walks of each pseudoscientific topic, we highlight some interesting findings.
For all topics, the amount of pseudoscientific content being suggested to the Pseudoscience profile after five hops is higher than the Science profile (see Fig.~\ref{fig:random_walks_pseudoscience_found}). 
In particular, the portion of unique pseudoscientific videos encountered by the Pseudoscience profile after five hops is $2.1\%$, $3.6\%$, $2.8\%$, and $7.1\%$ for COVID-19, Anti-vaccination, Anti-mask, and Flat Earth,  respectively, while for the Science profile, it is $0.8\%$, $1.9\%$, $0.0\%$, and $3.1\%$.
We also validate the statistical significance of the differences in the portion of pseudoscientific content suggested to the Pseudoscience profile compared to the Science profile for Anti-vaccination, Anti-mask, and Flat Earth, via the Fisher's Exact test ($p<0.05$).

Lastly, we find that for more traditional pseudoscientific topics like Flat Earth, YouTube suggests more pseudoscientific content to all types of users, except the YouTube Data API, compared to the other three more recent pseudoscientific topics. 
Using Fisher's exact test, we confirm that this difference between Flat Earth and COVID-19 is statistically significant for all types of users ($p<0.05$), 
while for Anti-mask this holds for the Science profile and the no profile (browser), and for Anti-vaccination this holds for the Pseudoscience profile and the no profile (browser).
This is another indication that YouTube has taken measures to counter the spread of pseudoscientific misinformation related to important topics like the COVID-19 pandemic. 

Overall, in most cases, the watch history of the user \emph{does} affect user recommendations and the amount of pseudoscientific content suggested by YouTube's algorithm.
This is also evident from the results of the random walks performed on the browser by the user with no profile. 
This profile does not maintain a watch history. 
It is recommended less pseudoscientific content than all the other profiles after five hops when starting from a video related to COVID-19 ($0.7\%$), and mainly to Anti-vaccination ($0.7\%$) and Flat earth ($3.1\%$).

Finally, we find a higher amount of pseudoscientific content in the random walks performed using the API than the random walks performed with the other non-logged-in user on the browser. 
In particular, the amount of unique pseudoscientific videos encountered by the YouTube Data API after five hops is $1.3\%$, $2.0\%$, $4.6\%$, and $4.2\%$ for COVID-19, Anti-vaccination, Anti-mask, and Flat earth, respectively, while, for the no profile (browser), it is $0.7\%$, $0.7\%$, $1.6\%$, and $3.1\%$.
However, this difference is not statistically significant and this indicates that the YouTube Data API results do not account for user personalization and the API does not maintain a watch history.
On the other hand, this difference may indicate that the YouTube Data API is more sensitive to item-to-item mapping~\cite{linden2003amazon} of the videos by the recommendation engine.

\subsection{Temporal Sensitivity}
Here, we investigate any variations in the results of our experiments over time either due to changes in the recommendation algorithm or to the effectiveness of the moderation strategies employed by YouTube.
Although we are not aware of any significant changes in the recommendation algorithm and YouTube has not officially announced any changes to its system, the company did announce during the COVID-19 pandemic that it will revert to human moderators to effectively tackle misinformation on its platform~\cite{ft2020humanannotators}.
Hence, to investigate the temporal sensitivity of our results, we perform the video recommendations experiment once again between April and May 2021 
using the same experiment setup and user profiles.
Figure~\ref{fig:random_walks_pseudoscience_found_second} plots the percentage of unique pseudoscientific videos that the random walker encounters at hop $k$ for each of the pseudoscientific topics explored.
Importantly, we find that, for all pseudoscientific topics, the Pseudoscience profile receives more pseudoscientific content that the Science/Pseudoscience and the Science profiles.
Overall, we make similar observations as with the results of the video recommendations experiment performed in December, 2020.
Next, we compare the results of each pseudoscientific topic with the respective results of the identical experiment performed in December. 
Importantly, we observe a slight decrease in the amount of pseudoscientific content being suggested to all user profiles for Flat Earth, while for the other topics the differences are negligible.
In general, we find that our results are not substantially affected by changes in the algorithm and data that may have occurred over the studied time period.
Nevertheless, we argue that the results of this work, which mostly derive from a single point in time are valuable.
This is because we mainly focus on timely pseudoscientific topics pertaining to the COVID-19 pandemic that are increasingly popular and of great societal interest.
The topics we analyze also allow us to:
1) evaluate the effectiveness of “novel” methodologies employed by YouTube to tackle misinformation around specific topics like COVID-19 [4]; 
2) investigate how YouTube responded to misinformation against crucial topics and whether the response was timely; and 
3) investigate the effectiveness of the usual mitigation strategies employed by YouTube when compared to other special mitigation strategies employed. 
The latter can be done by comparing the percentage of pseudoscientific content observed for traditional topics (i.e., Flat Earth) to the one observed for timely topics like COVID-19.

 \subsection{Take Aways}
 \label{subsec:experiments_take_aways}
We now summarize the main findings of our experiments.
Table~\ref{tab:experiments_pseudoscience_videos_found_details} reports the percentage of unique pseudoscientific videos appearing on the YouTube homepage, search results, and the video recommendations section for each user profile out of all the unique videos encountered by each user profile in each experiment.

The highest percentage of pseudoscientific videos occurs in the search results. %
That experiment shows that, for all pseudoscientific topics except COVID-19, the Pseudoscience and the Science/Pseudoscience profiles encounter more pseudoscientific content when searching for these topics than the Science profile.
For COVID-19, none of the profiles see any pseudoscientific content. 
When it comes to recommendations, in all the random walks (except Anti-mask), the Pseudoscience profile gets more pseudoscientific content than all the other profiles. 
For Anti-mask, we find a higher proportion of pseudoscientific content using the Data API.

Overall, the main findings of our analysis are:
\begin{compactenum}
\item The watch history of the user substantially affects what videos are suggested to the user.

\item It is more likely to encounter pseudoscientific videos in the search results (i.e., when searching for a specific topic) than in the video recommendations section or the homepage of a user, except in the case of the COVID-19 topic. 

\item For ``traditional'' pseudoscience topics (e.g., Flat Earth), there is a higher rate of recommended pseudoscientific content than for more emerging/controversial topics like COVID-19, anti-vaccination, and anti-mask. For COVID-19, we find an even smaller amount of pseudoscientific content being suggested, which may result from measures YouTube took to mitigate misinformation concerning the COVID-19 pandemic.

\item Although YouTube seems to tackle COVID-19 related misinformation in its search results, all profiles used in our experiments still receive recommendations to questionable content related to the pandemic.
\item The difference between the results of the YouTube Data API and the no profile (browser) is statistically insignificant; this indicates that recommendations returned using the API are not subject to personalization.This finding can be helpful to other researchers that use YouTube's Data API for their experiments.
\end{compactenum}

\section{Related Work}
\label{sec:related_work}

This section reviews prior work on pseudoscience, misinformation, and other malicious activity on YouTube, the recommendation algorithm, and user personalization.%

\descr{Pseudoscience and Misinformation.}
The scientific community has extensively studied the phenomenon of misinformation and the credibility issues of online content~\cite{kumar2018false}. %
The majority of previous work focuses on analyzing misinformation and pseudoscientific content on other social networks~\cite{rajdev2015fake,johnson2020online},
although some study specific misinformative and conspiratorial topics on YouTube.

For instance, Li et al.~\cite{li2020youtube} study misinformation related to the COVID-19 pandemic on YouTube;
they search YouTube using the terms ``coronavirus'' and ``COVID-19,'' and analyze the top 75 viewed videos from each search term, finding $27.5\%$ of them to be misinformation.
Donzelli et al.~\cite{donzelli2018misinformation} focus on misinformation surrounding vaccines that supposedly cause autism by performing a quantitative analysis of YouTube videos. 
Landrum et al.~\cite{landrum2019differential} investigate how users with varying science comprehension and attitude towards conspiracies are susceptible to Flat Earth arguments on YouTube. 
Faddoul et al.~\cite{faddoul2020longitudinal} develop a classifier to detect conspiratorial videos on YouTube and use it to perform a longitudinal analysis of conspiracy videos emulating YouTube's autoplay feature, without user personalization.
Hou et al.~\cite{hou2019towards} focus on the detection of misinformative videos related to prostate cancer. %
Then, they use linguistic, acoustic, and user engagement features and they develop classifiers that can detect misinformative videos with $74\%$ accuracy.
Serrano et al.~\cite{serrano2020nlp} focus on the detection of COVID-19 misinformation videos on YouTube and they propose an NLP-based classifier that is based on the comments and the title of a video and can detect COVID-19 misinformative videos with $89.4\%$.
Although this classifier has higher accuracy than our classifier, the fact that it is mainly based on the comments of a video is problematic; a malicious actor can manipulate the comments of a video while maintaining its content intact.
Overall, our work extends prior research as we focus on multiple health-related and other traditional misinformation topics on YouTube.  
We present a classifier and a novel methodology which allow us to assess the effects of a user's watch history on YouTube's pseudoscientific recommendations in multiple parts of the platform.  

\descr{Malicious activity on YouTube.}
A substantial body of work focuses on detecting and studying malicious content on YouTube.
Jiang et al.~\cite{jiang2019bias} investigate how channel partisanship affects comment moderation.
Zannettou et al.~\cite{zannettou2018good} propose a deep learning classifier for identifying videos on YouTube that use manipulative techniques to increase their views, i.e., clickbait. 
Mariconti et al.~\cite{enrico2019cscw} build a classifier to predict, at upload time, whether a YouTube video will be ``raided'' by hateful users. 

\descr{YouTube's Recommendation Algorithm and Audits.}
Zhao et al.~\cite{zhao2019recommending} introduce a large-scale ranking system for YouTube recommendations, which
ranks the candidate recommendations of a given video, taking into account user engagement and satisfaction metrics (e.g., video likes).
Next, Ribeiro et al.~\cite{ribeiro2020auditing} perform a large-scale audit of user radicalization on YouTube: they analyze videos from Intellectual Dark Web, Alt-lite, and Alt-right channels, showing that they increasingly share the same user base. 
Papadamou et al.~\cite{papadamou2020disturbed} focus on detecting disturbing videos on YouTube targeting young children finding that young children are likely to encounter disturbing videos when they randomly browse the platform.

\descr{User Personalization.}
Most of the work on user personalization focuses on Web search engines and is motivated by the concerns around the Filter Bubble effect~\cite{pariser2011filter}.
Hannak et al.~\cite{hannak2013measuring} propose a methodology for measuring personalization in Web search results.
Robertson et al.~\cite{robertson2018auditing} focus on the personalization and composition of politically-related search engine results, and they propose a methodology for auditing Google Search. %
Le et al.~\cite{le2019measuring} investigate whether politically oriented Google news search results are personalized based on the user's browsing history; using a ``sock puppet'' audit system, they find significant personalization, which tends to reinforce the presumed partisanship of a user.
St{\"o}cker et al.\cite{stocker2020riding} analyze the effect of extreme recommendations on YouTube, finding that YouTube's auto-play feature is problematic.
\revisionfinal{
Finally, Hussein et al.~\cite{hussein2020measuring} focus on measuring misinformation on YouTube search results and the video recommendations section considering five popular conspiratorial topics.
More precisely, they create user profiles with distinct demographics and watch history and use them to investigate the effects of these personalization attributes on the amount of misinformation in YouTube search results and video recommendations.
}

\revisionfinal{
Taking cues from~\cite{hussein2020measuring}, we complement their work by providing valuable additional insights and findings.
Unlike Hussein et al.~\cite{hussein2020measuring}, we measure what videos a user who follows YouTube's recommendations encounters, while they measure the recommendations suggested to a user who watches only a curated subset of the videos returned by a set of search queries.
In particular, we emulate the behavior of users with distinct and already established watch histories who start by watching a single video (as returned by the initial search query) and subsequently watch videos suggested by YouTube's recommendation algorithm after each view.
}

\revisionfinal{
Moreover, we devise a novel algorithm that allows us to study the impact of the length of the watch history of a user in the amount of personalization.
Regarding the topics that we analyze, we acknowledge that two of them are similar to two of the topics analyzed by~\cite{hussein2020measuring} (``Anti-vaccination'' and ``Flat Earth''). 
However, we mainly focus on multiple health-related pseudoscientific topics on YouTube pertaining to the COVID-19 pandemic. 
Last, we also investigate the temporal sensitivity of our results which derive from a single point in time, finding that they are representative.
}
\revcomment{(Comment: MT1)}

\section{Discussion \& Conclusion}
\label{sec:discussion_conclusion}
In this work, we studied pseudoscientific content on the YouTube platform. 
We collected a dataset of 6.6K YouTube videos, and by using crowdsourcing, we annotated them according to whether or not they include pseudoscientific content.
We then trained a deep learning classifier to detect pseudoscientific videos. 
We used the classifier to perform experiments assessing the prevalence of pseudoscientific content on various parts of the platform while accounting for the effects of the user's watch history.
To do so, we crafted a set of accounts with different watch histories.

\descr{Main Results.} Overall, we found that the user's watch history does substantially affect future user recommendations by YouTube's algorithm.
This should be taken into consideration by research communities aiming to audit the recommendation algorithm and understand how it drives users' content consumption patterns. 
We also found that YouTube search results are more likely to return pseudoscientific content than other parts of the platform like the video recommendations section or a user's homepage.
However, we also observed a non-negligible number of pseudoscientific videos on both the video recommendations section and the users' homepage.
By investigating the differences across multiple pseudoscientific topics, we showed that the recommendation algorithm is more likely to recommend pseudoscientific content from traditional pseudoscience topics, e.g., Flat Earth, compared to more controversial topics like COVID-19.
This likely indicates that YouTube takes measures to counter the spread of harmful information related to critical and emerging topics like the COVID-19 pandemic. 
However, achieving this in a proactive and timely manner across topics remains a challenge.

\descr{Looking Forward.} %
The relatively low agreement score of our crowdsourced annotation %
points to the difficulty in objectively identifying whether a video is pseudoscientific or not and also confirms that it is not easy to automate the discovery of misinformation.
Hence, we believe that the most proper way for YouTube to cope with misinformation on the platform effectively is to use deep learning models that signal potential pseudoscientific videos to human annotators who examine the videos and make the final decision.

Our work provides insights on pseudoscientific videos on YouTube and provides a set of resources to the research community (we will make the dataset, the classifier, and all the source code of our experiments publicly available). 
In particular, the ability to run this kind of experiments while taking into account users' viewing history will be beneficial to researchers focusing on demystifying YouTube's recommendation algorithm---irrespective of the topic of interest.
In other words, our methodology and codebase are generic and can be used to study other topics besides pseudoscience, e.g., additional conspiracy theories.

\descr{Limitations.} 
Naturally, our work is not without limitations.
First, we use crowdworkers who are unlikely to have any expertise in identifying pseudoscientific content. 
Hence, a small percentage of the annotated videos may be misclassified. 
We mitigated this issue by not including annotators with low accuracy on a classification task performed on a test dataset and annotating each video based on the majority agreement.
We also evaluated our crowdsourced annotation’s performance by manually reviewing a randomly selected set of videos from our ground-truth dataset, yielding $0.92$ precision, $0.91$ recall, and $0.92$ F1 score.
Second, our ground-truth dataset is relatively small for such a subjective classification task.
Nonetheless, the classifier provides a meaningful signal, which, supported by manual review, allows us to assess YouTube's recommendation algorithm's behavior with respect to pseudoscientific content.
Third, there might be videos in our experiments that are pseudoscientific and have been classified as ``Other.''
Hence, to verify our results' accuracy, we manually reviewed a random sample ($10\%$) of the videos encountered during our experiments and classified them as ``Other,'' finding that $98\%$ of them were correctly classified.
\revisionfinal{
Fourth, in our experiments we always watch $50\%$ of the total duration of a video, which is limited compared to a length-calibrated average watch percentage~\cite{wu2018beyond}.
However, to calculate this percentage we need the total watch time of the video. 
Unfortunately, this information for the type of videos that we analyze is not available anymore through the YouTube Data API.
}
\revcomment{(Comment: R.3.4)}
Finally, as for user personalization, we only work with watch history, which is a fraction of YouTube's signals for user personalization.

\descr{Future Work.}
A more comprehensive user personalization methodology to account for factors outside of watch history, such as account characteristics and user engagement, is a clear direction for future research. 
We also plan to extend our framework taking into account the ranking of the videos in the various parts of the YouTube platform.
Last, we plan to conduct studies to understand how people engage, share, and view pseudoscientific content on other social media platforms, including Twitter and Facebook.

\descr{Acknowledgments.} This project has received funding from the European Union's Horizon 2020 Research and Innovation program under the CONCORDIA project (Grant Agreement No. 830927),
and from the Innovation and Networks Executive Agency (INEA) under the CYberSafety II project (Grant Agreement No. 1614254).
This work reflects only the authors' views; the funding agencies are not responsible for any use that may be made of the information it contains.

\small
\bibliographystyle{abbrv}
%\bibliography{references}

\end{document}